\documentclass[12pt,a4paper]{article}

\usepackage[margin=2cm]{geometry}
\usepackage[hidelinks]{hyperref}
\usepackage{graphicx}
\usepackage{amssymb, amsmath, amsfonts, bm, mathrsfs,amsthm}
\usepackage{siunitx}
\usepackage[utf8]{inputenc}
\usepackage{cleveref}

\newcommand{\E}[0]{\mathcal{E}}
\renewcommand{\O}[0]{\mathcal{O}}

\newcommand{\eps}[0]{\varepsilon}

\newcommand{\notvc}[1]{#1}

\newcommand{\U}[0]{\mathcal{U}}

\newcommand{\grad}[0]{\nabla}
\newcommand{\lapof}[0]{\nabla^2}
\newcommand{\divof}[0]{\nabla \cdot}
\newcommand{\abs}[1]{\left|#1\right|}
\newcommand{\br}[1]{\left(#1\right)}

\newcommand{\pd}[2]{\frac{\partial #1}{\partial #2}}
\newcommand{\ehat}[0]{\hat{e}}
\renewcommand{\O}[0]{\mathcal{O}}
\newcommand{\nhat}[0]{\widehat{n}}
\newcommand{\that}[0]{\widehat{t}}
\newcommand{\Kbar}[0]{\overline{K}}
\newcommand{\Kabs}[0]{\abs{K}}
\newcommand{\Khat}[0]{\widehat{K}}

\newcommand{\Jhat}[0]{\widehat{J}}
\newcommand{\sgrad}[0]{\nabla_\bot}
\newcommand{\grads}[0]{\nabla^\bot}
\newcommand{\sdiv}[0]{\nabla_\bot \cdot}
\newcommand{\divs}[0]{\nabla^\bot \cdot}

\newcommand{\ds}[0]{\partial_\sigma}
\newcommand{\dx}[0]{\partial_\xi}

\newcommand{\ohat}[0]{\hat{\omega}}
\newcommand{\J}[0]{\mathcal{J}}
\newcommand{\K}[0]{\mathcal{K}}

\title{Improved phase-field models of melting and dissolution in multi-component flows}

\author{Eric W. Hester \footnote{University of Sydney School of Mathematics and Statistics, Sydney, NSW 2006, Australia (\href{mailto:eric.hester@sydney.edu.au}{eric.hester@sydney.edu.au}).
	 }
\and Louis-Alexandre Couston \thanks{British Antarctic Survey, Cambridge, CB3 0ET, UK} \thanks{Department of Applied Mathematics and Theoretical Physics, University of Cambridge, Cambridge, UK}
\and Benjamin Favier \thanks{Aix Marseille Univ, CNRS, Centrale Marseille, IRPHE, Marseille, France}
\and Keaton J. Burns \thanks{Massachusetts Institute of Technology Department of Mathematics, Cambridge, MA 02139, USA}
\thanks{Center for Computational Astrophysics, Flatiron Institute, Simons Foundation, New York, NY 10010, USA}
\and Geoffrey M. Vasil \footnotemark[1]}

\begin{document}

\maketitle

\begin{abstract}
We develop and analyse the first second-order phase-field model to combine melting and dissolution in multi-component flows.
This provides a simple and accurate way to simulate challenging phase-change problems in existing codes.
Phase-field models simplify computation by describing separate regions using a smoothed phase field.
The phase field eliminates the need for complicated discretisations that track the moving phase boundary.
However standard phase-field models are only first-order accurate.
They often incur an error proportional to the thickness of the diffuse interface.
We eliminate this dominant error by developing a general framework for asymptotic analysis of diffuse-interface methods in arbitrary geometries.
With this framework we can consistently unify previous second-order phase-field models of melting and dissolution and the volume-penalty method for fluid-solid interaction.
We finally validate second-order convergence of our model in two comprehensive benchmark problems using the open-source spectral code Dedalus.
\end{abstract}

\section{Introduction}
\label{sec:intro}
Many scientific and industrial questions involve fluid flows coupled with phase changes; including sea-ice formation \cite{EpsteinComplexFreezingMeltingInterfaces1983}, semiconductor crystal manufacture 
\cite{
GlicksmanInteractionFlowsCrystalMelt1986}, 
binary alloy solidification \cite{CahnFreeEnergyNonuniform1958},
and geophysical mantle dynamics \cite{HuppertGeologicalFluidMechanics2002}.
Multi-phase interaction combines the challenges of nonlinear multi-component convection \cite{
TurnerMulticomponentConvection1985} and evolution of phase boundaries \cite{KnoblochProblemsTimeVaryingDomains2015}, creating entirely new effects.
Quantifying this complexity demands appropriate mathematical tools.

Moving boundary problems are the standard method to model phase change phenomena.
Separate partial differential equations (PDEs) exist in the liquid and solid regions and moving boundary conditions are applied at the interface (see \cref{fig:sharp-vs-phase-diagram} $(a)$).
A dynamically shifting interface means that the boundary conditions form an essential (often nonlinear) part of the solution \cite{WorsterSolidificationFluids2002}.
Moving boundaries present many challenges, complicating 
numerical algorithms \cite{DoneaArbitraryLagrangianEulerian2004} 
and mathematical proofs \cite{
	HadzicWellposednessClassicalStefan2017}. 

As a possible remedy, it is useful to recall that boundary conditions are a mathematical abstraction; 
they result from limiting cases of rapid transitions in material properties.
There is a long history of reinterpreting discontinuous boundary conditions as smoothed phenomena.
Where Gibbs treated capillarity with infinitesimal surfaces  \cite{GibbsEquilibriumHeterogeneousSubstances1878},
Van der Waals understood the importance of smoothness at phase boundaries \cite{VanDerWaalsThermodynamicTheoryCapillarity1979}.
Where Stefan treated solid-liquid phase boundaries as discontinuous  \cite{StefanUeberTheorieEisbildung1891},
Cahn and Hilliard modelled phase separation as smoothed \cite{CahnFreeEnergyNonuniform1958}. 
Readopting a physics-based viewpoint of boundary conditions allows new possible techniques for addressing complex multi-phase problems. 
As well as providing a firmer mathematical and physical foundation, smoothed models also simplify numerical implementations by removing the need to track the infinitesimal boundary.

This paper focusses on phase-field models, one of the foremost examples of this smoothed approach.
Phase-field models represent distinct phases using a single smoothed \emph{phase field} $\phi$, illustrated in \cref{fig:sharp-vs-phase-diagram} $(b)$ \cite{
BoettingerPhaseFieldSimulationSolidification2002,
SteinbachPhasefieldModelsMaterials2009}.
The evolution of the phases is then determined by a single set of equations that apply over the entire domain.
Many other methods also model phase changes, such as enthalpy methods \cite{VollerFixedGridTechniques1990,UlvrovaNumericalModellingConvection2012}, level set methods \cite{
OsherLevelSetMethods2001,
ChenSimpleLevelSet1997}, diffuse-domain approaches \cite{LiSolvingPDEsComplex2009,AlandTwophaseFlowComplex2010}, or some immersed-boundary methods \cite{MacHuangStableAccurateScheme2020}.
Yet phase-field models stand out for combining several key benefits:
\begin{itemize}
	\item They are {physically motivated}, introduced by Fix \cite{FixPhaseFieldMethods1982} and Langer \cite{LangerModelsPatternFormation1986} to model free energy near phase boundaries (following from Hohenburg and Halperin's model C \cite{HohenbergTheoryDynamicCritical1977}).
	\item They {generalise canonical models} of phase separation, reducing to Allen-Cahn and Hele-Shaw flow (among others) in various asymptotic limits \cite{CaginalpStefanHeleShawtype1989,CaginalpConvergencePhaseField1998}.
	\item They are {easily extensible} to more general systems, 
	including two-component alloys \cite{WheelerPhasefieldModelIsothermal1992,
	BiPhaseFieldModel1998,
	KarmaPhaseFieldFormulationQuantitative2001,
	RamirezPhasefieldModelingBinary2004},
	convection \cite{BeckermannModelingMeltConvection1999,
	AndersonPhasefieldModelSolidification2000},
	or multi-phase flows \cite{AlandTwophaseFlowComplex2010,
	MokbelPhasefieldModelFluid2018}. 
	\item 	They can be made {thermodynamically consistent} \cite{PenroseThermodynamicallyConsistentModels1990,
	WangThermodynamicallyconsistentPhasefieldModels1993,
	McFaddenThinInterfaceAsymptotics2000,
	OhnoVariationalformulationnumerical2016,
	BolladaBracketformalismapplied2017}.
	\item They are mathematically rigorous, with well-posedness and convergence results \cite{CaginalpAnalysisPhaseField1986,
CaginalpDynamicsLayeredInterfaces1988}.
	\item They are simple to simulate as they avoid explicit tracking of the interface \cite{FixPhaseFieldMethods1982,
LinNumericalAnalysisPhase1988,
WheelerComputationdendritesusing1993,
FabbriPhaseFieldMethodSharpInterface1997,
TongPhasefieldSimulationsDendritic2001,
LuThreedimensionalphasefieldsimulations2005,
JokisaariBenchmarkproblemsnumerical2017,
FavierRayleighBenardconvection2019,
PurseedBistabilityRayleighBenardConvection2020,
CoustonIceMeltingTurbulent2020}.
\end{itemize}
We emphasise this last point.
The simulation of moving boundary problems requires specialised algorithms designed to track and apply boundary conditions at the interface.
These algorithms can be difficult or impossible to implement in existing codes.
For example, spectral methods are popular for their efficiency,
but cannot easily handle non-trivial geometries or topologies.
Phase-field models (and other diffuse-interface methods) alleviate these difficulties by removing boundary conditions from the problem formulation.
By replacing boundary conditions with simple source terms, they can be implemented in general codes for little effort.
More general effects can be modelled by changing source terms, 
as opposed to developing and integrating new algorithms into the codebase.
Phase-field models extend the range of phenomena that existing codes can simulate, 
and accelerate the development of codes to study new scientific problems.

However, phase-field models possess one important drawback for simulation: they must resolve the diffuse interface.
For small-scale simulations this is feasible.
But there is a vast disparity between the microscopic scale of the smoothed interface and the macroscopic scale of interest in most problems.
This disparity is what makes discontinuous boundary conditions appropriate models in most circumstances.
Throughout this paper, we denote this ratio by $\eps$
    \begin{align}
    \eps  = \frac{\text{microscopic interface width}{}}{\text{bulk system size}}.
    \end{align}
Numerically feasible values of $\eps$ are orders of magnitude larger than reality.
Straightforward analysis implies a commensurate $\O(\eps)$ error with respect to the limiting boundary conditions.

The only way to perform accurate phase-field simulations with achievable values of $\eps$ is to accelerate the convergence of the model itself. 
This can be done through second-order asymptotic analysis in the limit $\eps \to 0$.
While the first order is sufficient to determine the limiting behaviour,
it is the second order that reveals the dominant error of the model.
It is then possible to find optimal prescriptions that cancel the dominant error and boost convergence from $\O(\eps)$ to $\O(\eps^2)$.

This strategy leads to various `quantitative' (i.e.~second-order) phase-field models,
beginning with a correction for arbitrary interface-kinetics in pure materials \cite{KarmaPhasefieldMethodComputationally1996,
KarmaQuantitativePhasefieldModeling1998}, 
and since extended to unequal diffusivities \cite{AlmgrenSecondOrderPhaseField1999,McFaddenThinInterfaceAsymptotics2000}, 
multiple components \cite{KarmaPhaseFieldFormulationQuantitative2001}, 
and the combination thereof \cite{GarckeSecondorderphase2006,OhnoQuantitativePhasefieldModeling2009}. 
An introduction to this asymptotic procedure can be found in \cite{FifeDynamicsInternalLayers1988}.
Despite much success, progress is difficult. 
Second-order asymptotic analysis has not yet ascertained a quantitative phase-field model of multi-component convection\footnote{
Shortly before submission we became aware of the work \cite{SubhedarDiffuseInterfaceModels2020}. In it, Subhedar et al.~perform second-order analysis of a phase-field model combining melting and advection.
Our work is more general as we also account for dissolution, give a more comprehensive analytical treatment, and use more challenging computational benchmarks.}.

This paper presents the first second-order phase-field model of buoyancy-forced convecting binary mixtures.
The model, given in \cref{sec:maths}, builds on first-order models of multicomponent convection \cite{BeckermannModelingMeltConvection1999},
second-order models of pure melts \cite{ChenRapidlyConvergingPhase2006}, 
the diffuse domain method for Robin boundary conditions \cite{KockelkorenComputationalApproachModeling2003},
and the smooth volume-penalty method for no-slip boundary conditions  \cite{HesterImprovingConvergenceVolume2019}.
We verify second-order convergence in \cref{sec:asymptotics} by developing a straightforward asymptotic procedure suitable for general equations and geometries in three dimensions.
This procedure allows us to consistently analyse and unify previous second-order phase-field and diffuse-interface methods.
We also implement this procedure in the symbolic computing language Mathematica.
For brevity we assume somewhat simplified thermodynamical properties in our model, such as uniform temperature diffusivity, negligible solute within the solid, and Boussinesq buoyancy.
Each assumption could be relaxed and analysed using the framework of \cref{sec:asymptotics}.
We finally validate the improved convergence in two comprehensive benchmark problems implemented in the Dedalus numerical code \cite{BurnsDedalusFlexibleFramework2020} in \cref{sec:experiments}.

\section{Models of melting in binary mixtures}
\label{sec:maths}
\subsection{Conventional moving boundary formulation}
\label{sec:sharp}
\begin{figure}[t]
        \centering
        \includegraphics[width=\linewidth]{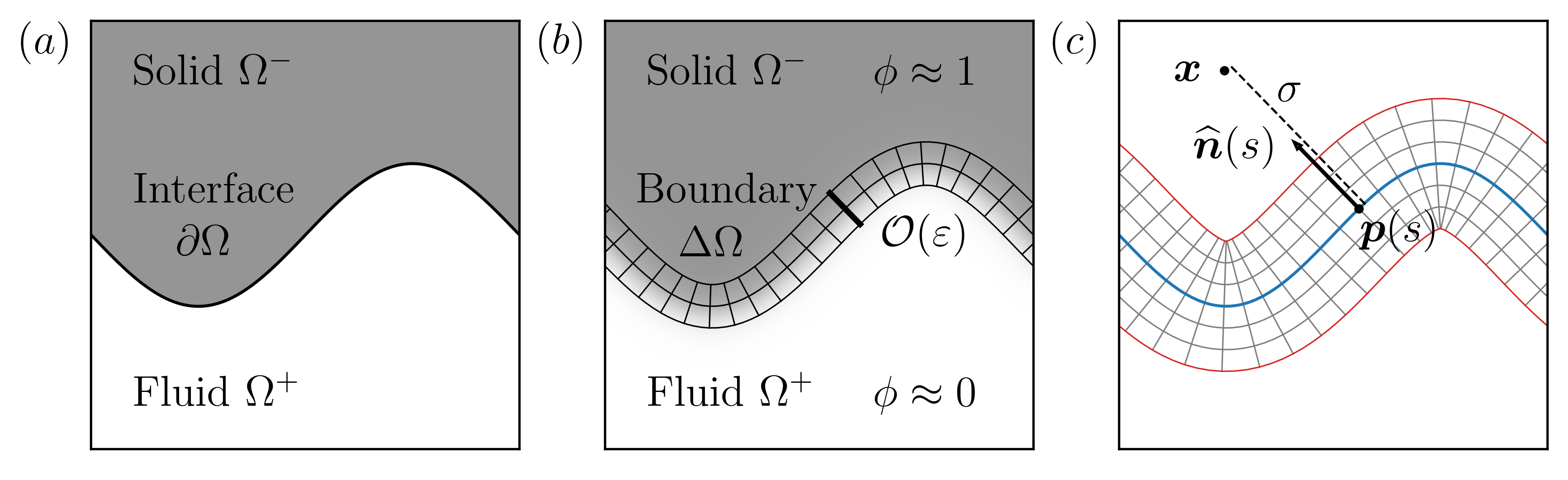}
        \caption{
        Figure $(a)$ illustrates a moving boundary formulation in which the domain is partitioned into the fluid $\Omega^+$, solid $\Omega^-$ and interface $\partial\Omega$.
        Figure $(b)$ show a phase-field approximation, where the phase $\phi$ smoothly varies between fluid and solid.
        The dashed line highlights the $\phi=1/2$ contour that approximates the true interface.
        We illustrate the asymptotic fluid $\Omega^+$, solid $\Omega^-$, and size $\O(\eps)$ boundary $\Delta\Omega$ regions of our analysis.
		Figure $(c)$ details the signed-distance coordinate system used to analyse the boundary region $\Delta\Omega$.
		Points on the interface $\notvc{p}(s)$ (blue) are parameterised by arbitrary surface coordinates $s$.
		A point off the boundary $x$ can be reached by moving a distance $\sigma$ in the normal direction $\notvc{\nhat}(s)$ from the closest point on the manifold $\notvc{p}(s)$. 
		Coordinate singularities (the corners of the red curves) will occur, but the coordinate system remains well-behaved in the interface region (fig.~$(b)$).
		The figure is in two dimensions for clarity, though the analysis of \cref{sec:asymptotics} is done in three dimensions.}
    \label{fig:sharp-vs-phase-diagram}
    \end{figure}

Melting in binary mixtures, such as ice in sea water, is often modelled as a moving boundary problem.
We partition the domain into fluid $\Omega^+$ and solid $\Omega^-$ regions, 
pose separate PDEs on each subdomain, 
and apply boundary conditions at the evolving interface $\partial \Omega$ (as in \cref{fig:sharp-vs-phase-diagram} (a)).

In the fluid, the temperature $T^+$ and dissolved solute concentration $C$ satisfy advection-diffusion equations, 
and the fluid velocity $\notvc{u}$ and pressure $p$ satisfy incompressible Navier-Stokes equations with Boussinesq buoyancy forcing $g \rho(T^+,C) \, \hat{z}$,
    \begin{align}
	\label{eq:sharp-pdes}
    \partial_t T^+ + \notvc{u}\cdot\nabla T^+ - \kappa \lapof T^+ &= 0, \nonumber\\
    \partial_t C + \notvc{u} \cdot\nabla C - \mu \lapof C &= 0, \nonumber\\
    \partial_t\notvc{u}+ \notvc{u} \cdot\nabla\notvc{u}- \nu \lapof\notvc{u}+ \nabla p + \frac{g\rho(T^+,C)}{\rho_0}\notvc{\hat{z}} &= 0,\nonumber\\
    \nabla \cdot\notvc{u} &= 0, \quad \text{in} \quad \Omega^+,
    \end{align}
where $\kappa,\mu,$ and $\nu$ are the thermal, solutal, and momentum diffusivity (each assumed constant), and $g$, $\rho$, and $\rho_0$ are the gravitational acceleration, the density of the fluid, and a reference density.
In the solid only the temperature $T^-$ is defined, which follows a diffusive equation,
    \begin{align}
    \partial_t T^- - \kappa \lapof T^- &= 0, \quad \text{in} \quad \Omega^-.
    \end{align}

We require several boundary conditions at the moving interface \cite{WorsterSolidificationFluids2002}.
The Gibbs-Thompson relation relates departure of thermosolutal equilibrium ($T+mC$, where $m$ is the liquidus slope) to a mean-curvature $\Kbar$ dependent surface energy, and kinetic undercooling proportional to the interfacial normal velocity $v$.
The Stefan condition expresses heat conservation, equating latent heat $L$ release with a discontinuity in temperature flux.
The Robin concentration condition ensures total solute conservation.
Zero velocity boundary conditions maintain mass conservation,	
    \begin{align}\label{eq:sharp-bcs}
    T + m C - \gamma \Kbar + \alpha v &= 0, &
    [\kappa \nhat \cdot \nabla T]^+_- + L v &= 0,&
    \mu \nhat \cdot \nabla C^+ + C^+ v&= 0,&
   \notvc{u}&= 0.
    \end{align}
This model involves several idealisations, namely Boussinesq buoyancy forcing, constant diffusivities and densities that are phase, temperature, and concentration invariant, and a linear liquidus relation.
We can include more general thermodynamic properties into our framework, but we continue with the current model as it captures many aspects of melting and dissolution in multi-component flows.
Kinetic undercooling is only relevant for rapidly solidifying supercooled liquids so we set $\alpha = 0$.
Note that the neglect of density change during melting means the fluid velocity is equal to zero at the moving interface.

\subsection{Phase-field model}
\label{sec:smooth}
Phase-field models represent an alternative approach that is physically motivated and simple to simulate.
They represent distinct phases with a smoothed \emph{phase field} $\phi$.
This field obeys an Allen-Cahn type equation which forces the phase to $\phi\approx 1$ in the solid and $\phi\approx 0$ in the fluid \cite{BeckermannModelingMeltConvection1999}.
The interface is represented implicitly by the level set $\phi = 1/2$.
The new equations augment are,
    \begin{align}
	\label{eq:phase-pdes}
    \partial_t T + \notvc{u} \cdot\nabla T - \kappa \lapof T &= L \partial_t \phi ,\nonumber\\
    \partial_t C + \notvc{u} \cdot\nabla C - \mu \lapof C &= \frac{ C\partial_t\phi  - \mu \nabla \phi \cdot \nabla C}{1-\phi+\delta},\nonumber\\
    \partial_t\notvc{u}+ \notvc{u} \cdot\nabla\notvc{u}- \nu \lapof\notvc{u}+ \nabla p + \frac{g\rho(T,C)}{\rho_0}\notvc{\hat{z}} &= - \frac{\nu}{(\beta\eps)^2} \phi \,\notvc{u},\nonumber\\
    \nabla \cdot\notvc{u}&= 0,\nonumber\\
    \eps\frac{5}{6}\frac{L}{\kappa}\partial_t\phi -\gamma \lapof \phi &= - \frac{1}{\eps^2}\phi(1-\phi)(\gamma (1 - 2\phi) +\eps(T + m C)). 
    \end{align}
    
The new source terms of \cref{eq:phase-pdes} can be understood heuristically.
At leading order, the phase-field equation develops a tanh-like profile around the interface with thickness $\eps$.
Beyond this distance the phase $\phi$ tends to its limiting values of zero in the fluid and one in the solid.
In the fluid \cref{eq:phase-pdes} reproduces \cref{eq:sharp-pdes}.
in the solid the advective and diffusive solute flux tend to zero (with $\delta \ll 1$ regularising the concentration equation for numerical stability),
and the velocity is damped by Darcy drag terms for porous media.
At next order thermosolutal forcing perturbs the interface to generate latent heat.
For a clear derivation of a similar first-order model see \cite{BeckermannModelingMeltConvection1999}.

Our goal is to surpass this approximate understanding and demonstrate improved convergence of this optimised phase-field model to the original moving boundary formulation.
We achieve this with an asymptotic analysis of \cref{eq:phase-pdes} as the interface length-scale $\eps$ tends to zero.
Proving these equations converge to the moving boundary formulation at $\O(\eps^2)$ in general geometries is nontrivial, 
but builds on second-order models of each individual boundary condition; 
a phase-field model which optimises the mobility term for zero interface kinetics \cite{ChenRapidlyConvergingPhase2006},
a concentration equation similar to \cite{BeckermannModelingMeltConvection1999} 
and the diffuse domain method for Robin boundary conditions \cite{KockelkorenComputationalApproachModeling2003},
and the smooth volume penalty method (which gives $\beta = 1.51044385$) \cite{HesterImprovingConvergenceVolume2019}.

\section{Analysis of the phase-field model}
\label{sec:asymptotics}

In order to understand the phase-field model and demonstrate second-order accuracy we use a multiple-scales matched-asymptotics framework, which we break into several modular steps:
\begin{enumerate}
	\item Partition the solid $\Omega^-$, fluid $\Omega^+$, and size $\O(\eps)$ boundary $\Delta \Omega$ regions  (\cref{fig:sharp-vs-phase-diagram} $(b)$).
	\item Adopt signed distance coordinates in the boundary region (\cref{sec:sdf}, \cref{fig:sharp-vs-phase-diagram} $(c)$).
	\item Rescale normal coordinate and operators by $\eps$ near the interface (\cref{sec:scaled-sdf}, \cref{fig:sharp-vs-phase-diagram} $(b)$).
	\item Expand the variables in an asymptotic power series in $\eps$ in each region (\cref{sec:power-series}).
	\item Connect regions with asymptotic matching conditions in the limit $\eps \to 0$ (\cref{sec:matching}).
	\item Iterate to solve the zeroth, first, and second order problems (\cref{sec:zeroth,sec:first,sec:second}).
\end{enumerate}
This procedure follows our previous analysis of the volume-penalty method \cite{HesterImprovingConvergenceVolume2019}.
The philosophy of our approach is to determine the evolution of the phase-field model up to and including $\O(\eps^2)$ in each region.
We show that the variables in the fluid and solid regions, as well as the location of the interface itself, evolve with only $\O(\eps^2)$ divergence from the moving boundary formulation.
Errors of $\O(\eps)$ in the temperature and tangential velocity do occur in the boundary region $\Delta \Omega$.
But this deviation is a necessary consequence of smoothly approximating discontinuous gradients across the interface, and is localised to the boundary region.
We thereby derive a second-order accurate phase-field model.
We now summarise the key components of this procedure\footnote{We also provide a Mathematica script that automates each step of this analysis at \href{https://github.com/ericwhester/phase-field-code}{github.com/ericwhester/phase-field-code}.}.

\subsection{Summary of asymptotic procedure}
\label{sec:framework}
\subsubsection{Signed-distance coordinate system}
\label{sec:sdf}
We build a simple orthogonal coordinate system in the boundary region $\Delta \Omega$ using the signed-distance function from the $\phi=1/2$ level set.
The signed distance $\sigma$ is the minimum distance of a point $x$ to the interface.
It follows that the point $x$ must lie in the direction of the unit normal vector $\nhat$ from the nearest point on the interface $p$, which we label with surface coordinates $s$,
    \begin{align}\label{eq:sdf}
    x = p(s) + \sigma\, \nhat(s).
    \end{align}
The surface coordinates induce a tangent vector basis $t_i$.
Given orthogonal surface coordinates, we also derive the dual vector basis $\nabla s_i$ and a unique orthonormal tangent basis $\that_i$,
    \begin{align}
    {t}_i &= \frac{\partial {p}}{\partial s_i}, &
    \that_i &= \frac{t_i}{|t_i|}, &
    \nabla s_i &= \frac{\that_i}{|t_i|}.
    \end{align}
We can then write the surface area measure $dA = |t_1||t_2|ds_1ds_2$ and surface gradient $\sgrad$,
    \begin{align*}
    \sgrad &= \nabla s_1  \frac{\partial}{\partial s_1} + \nabla s_2  \frac{\partial}{\partial s_2} = \that_1 \nabla_1 + \that_2 \nabla_2.
    \end{align*}
It is not difficult to show the normal $\nhat$ is everywhere equal to the gradient of the signed distance.
The gradient of the normal is therefore symmetric and diagonalisable.
The eigenvectors are the principal directions of curvature (which must align with orthogonal surface coordinates), 
and the eigenvalues are the principal curvatures $\kappa_i$,
    \begin{align}
    \nhat &= \nabla \sigma, &
    \nabla \nhat &= - \kappa_1 \that_1 \that_1 - \kappa_2 \that_2 \that_2 = - K.
    \end{align}
We use this orthonormal frame to describe all geometric quantities near the interface.
We can then express the gradient using the surface and normal derivatives and the scale tensor $J$
    \begin{align}
    \nabla &= \nhat \ds + J^{-1} \cdot \nabla_\bot, \quad \text{where} \quad     J = I - \sigma K.
    \end{align}
It is straightforward to derive the remaining vector calculus operators from the gradient, which we list in \cref{app:sdf}.
Finally, the Cartesian partial time derivative $\partial_t$ can be rewritten in a moving coordinate system using the signed-distance partial time derivative $\partial_\tau$,
	\begin{align}
    \partial_t &= \partial_\tau  - v \partial_\sigma + \sigma \sgrad v \cdot J^{-1} \cdot \sgrad.
    \end{align}
    
\subsubsection{Rescaling interfacial coordinates}
\label{sec:scaled-sdf}
We analyse the size $\eps$ interfacial region using the rescaled coordinate $\xi$ and derivative $\partial_\xi$,
    \begin{align}
    \sigma &= \eps \xi, &
    \ds &= \frac{1}{\eps}\dx.
    \end{align}
This rescales vector calculus operators (listed in 
\cref{app:sdf}) through scale factors of the gradient
    \begin{align}
	    J &= I - \eps\xi K, & 
    J^{-1} &= \sum\nolimits_{k=0}^\infty \eps^k \xi^k K^k.
   	\end{align}
\subsubsection{Variable expansions with formal power series}
\label{sec:power-series}

After splitting the domain into the fluid $\Omega^+$, solid $\Omega^-$, and interfacial $\Delta \Omega$ regions, 
each variable $f$ in each region (fluid $f^+$, solid $f^-$, and interfacial $f$) is expressed as a power series in $\eps$,
	\begin{align}
	f^+(x,t) &=    \sum_{k=0}^\infty \eps^k f^+_k(x,t), &
	f^-(x,t) &=    \sum_{k=0}^\infty \eps^k f^-_k(x,t), &
	f(\xi,s,\tau) &= \sum_{k=0}^\infty \eps^k f_k(\xi,s,\tau).
	\end{align}
We do so for the temperature $T$, solute concentration $C$, fluid velocity $u = u_\sigma \nhat + u_\bot$, pressure $p$, phase field $\phi$, and interface velocity $v\, \nhat$ (which does not depend on $\xi$).
We substitute these series into the hierarchy of equations generated by \cref{sec:scaled-sdf} to derive a system of equations at each order of $\eps$.
Solving each order requires a matching procedure between adjacent regions.

\subsubsection{Asymptotic matching}
\label{sec:matching}
To ensure agreement between different regions we specify asymptotic matching boundary conditions.
This subtle notion requires asymptotic agreement in intermediate zones $\xi \sim \eps^{-1/2}$ in the limit that $\eps\to 0$.
We can then let $\xi$ approach infinity for the inner problem without encountering coordinate singularities (provided $ \eps \ll \min_i|\kappa_i^{-1}|$), 
and let $\sigma$ approach zero for the outer problem without entering the interfacial region.
That is, for any variable $f$, we require
    \begin{align}
	\lim_{\eps\to 0} f(\pm\eps^{-1/2}\xi,s,t) &\sim \lim_{\eps\to 0} f^{\pm}(p(s,t)\pm\eps^{+1/2}\,\xi \,\nhat(s,t),t).
    \end{align}
Each variable is already expressed as an asymptotic series.
Each series term in the outer variables can be further expanded as a Taylor series about the interface at $\phi = 1/2$ using the signed distance $\sigma$.
The matching conditions at each order of $\eps$ then simplify to
    \begin{align}
    \lim_{\xi\to\pm\infty} f(\xi) = \lim_{\xi\to\pm\infty} \sum_{k=0} \eps^k f_k 
    \sim \lim_{\xi\to\pm\infty}  \sum_{k=0}\eps^k\br{\sum_{\ell=0}^k\frac{\xi^{\ell}}{\ell!} \ds^\ell f^{\pm}_{k-\ell}|_{\sigma=0}}.
    \end{align}
\begin{subequations}
Equipped with our multiple-scales matched-asymptotics procedure, we can now verify second-order convergence of our phase-field model in general smooth geometries.
\end{subequations}

\subsection{Zeroth order}
\label{sec:zeroth}
At leading order the phase-field equation reduces to the condition
	$\phi_0^\pm(1-\phi_0^\pm)(1-2\phi_0^\pm) = 0$.
We define the solid by $\phi^-_0 = 1$, the liquid by $\phi_0^+ = 0$, with the boundary region separating them.

\emph{Fluid problem} --- Noting that the phase is zero in the fluid, we recover the desired sharp interface equations for the remaining leading order variables,
    \begin{align*}
    \partial_t T^+_0 + u^+_0\cdot\nabla T^+_0 - \kappa \lapof T^+_0 = 0, \qquad
    \partial_t C^+_0 + u^+_0\cdot\nabla C^+_0 - \mu \lapof C^+_0 = 0,\\
	    \partial_t u^+_0 + u^+_0\cdot\nabla u^+_0 - \nu \lapof u^+_0 + \nabla p^+_0 - B(T^+_0- N C^+_0) \hat{z}= 0, \qquad
	    \nabla \cdot u^+_0 = 0.
    \end{align*}
For brevity we approximate the buoyancy relationship as being linear in temperature and concentration, with proportionality constants $B$ and $-NB$ respectively.
This simplification does not affect the convergence results.
The solutions may also require external boundary conditions to complete the system.
    
\emph{Solid problem} --- Within the solid, we reproduce the diffusion equation for the temperature, and zero velocity in the solid.
The divergence of the momentum equation reveals a Poisson equation for the pressure.
The concentration equation depends sensitively on the decay of the phase to zero,
but in the region $1-\phi \ll \delta$ the concentration forcing terms vanish, giving
	\begin{align*}
    \partial_t T^-_0 - \kappa \lapof T^-_0 &= 0, &
    \partial_t C^-_0 - \mu \lapof C^-_0 &= 0, &
	u^-_0 &= 0, &
	\nabla^2 p^-_0 &= B\partial_z (T^-_0 - N C^-_0).
    \end{align*}
Similarly, internal ice boundary conditions may be necessary.

\emph{Boundary problem} --- 
The boundary region is defined relative to the interface $\phi = 1/2$.
To hold true for all $\eps$ when expanded into its power series, this implies that 
	\begin{align*}
	\phi_0(\xi = 0) = 1/2 \quad \text{and} \quad \phi_k(\xi=0) = 0 \quad \text{for all } k>0.
	\end{align*}
In this region the phase-field equation balances the diffusion and reaction terms.
The limiting boundary conditions $\phi_0(\xi\to+\infty) = 0$ and $\phi_0(\xi\to-\infty) = 1$ imply a tanh profile for $\phi_0$,
    \begin{align*}
    \dx^2\phi_0 &= \phi_0(1-\phi_0)(1-2\phi_0), & 
    \lim_{\xi\to\pm\infty}\phi_0(\xi) &\sim \phi^\pm_0|_{\sigma=0}, &
    \implies & &
	\phi_0(\xi) &= \frac{1}{2}\br{1 - \tanh\frac{\xi}{2}}.
	\end{align*}
Matching the inner heat equation to the outer temperature requires $T_0$ to be constant in $\xi$,
    \begin{align*}
    \dx^2 T_0 &= 0,  &
    \lim_{\xi\to\pm\infty}T_0(\xi) &\sim T^\pm_0|_{\sigma=0}, &
    \implies & &
    T_0 &= T_0^\pm|_{\sigma=0}.
    \end{align*}
The identity $	\dx \phi_0 = - \phi_0 (1-\phi_0)$ simplifies the concentration equation.
The limiting behaviour of the operator implies exponential growth of the kernel into the solid.
This is unphysical, implying that the operand $\dx C_0$ is zero.
Matching to the outer variables requires a constant concentration,
    \begin{align*}
    (1-\phi_0)(\dx + \phi_0)\dx C_0 &= 0, & 
    \lim_{\xi\to\pm\infty}C_0(\xi) &\sim C^\pm_0|_{\sigma=0} &
    \implies & &
	C_0 &= C_0^\pm|_{\sigma=0}.
	\end{align*}
The divergence constraint and matching to the solid implies zero normal velocity,
	\begin{align*}
	\dx u_{\sigma 0} &= 0, &
    \lim_{\xi\to\pm\infty}u_{\sigma 0}(\xi) &\sim u^\pm_{\sigma 0}|_{\sigma=0}, &
	\implies & &
	u_{\sigma 0} &= u^\pm_{\sigma 0}|_{\sigma=0} = 0.
	\end{align*}
The tangential momentum equation balances diffusion with damping.
The kernel is spanned by an unphysical solution that grows exponentially into the solid,
and a physical solution that decays exponentially into the solid and is affine in the fluid.
Calibrating $\beta$ \cite{HesterImprovingConvergenceVolume2019} allows linear (rather than affine) behaviour in the fluid.
Matching to the fluid implies $u_{\bot 0}$ is zero,
	\begin{align*}
    \br{\dx^2 - \frac{1}{\beta^2}\phi_0} u_{\bot0} &= 0, & 
    \lim_{\xi\to\pm\infty}u_{\bot 0}(\xi) &\sim u^\pm_{\bot 0}|_{\sigma=0}, &
    \implies & &
	u_{\bot 0} = u^\pm_{\bot 0}|_{\sigma=0} &= 0.
	\end{align*}
In summary, the zeroth order asymptotics shows that the phase field has a tanh profile near the boundary, and that the leading order outer velocities satisfy no-slip boundary conditions.
The next order problem reproduces the remaining boundary conditions.

\subsection{First order}
\label{sec:first}
The first-order perturbation of the phase-field equation away from the interface takes the form
    \begin{align*}
    (1 - 6 \phi_0^\pm + {6\phi_0^\pm}^2)\phi_1^\pm = (T_0^\pm+ m C_0^\pm) \phi_0^\pm(1-\phi_0^\pm).
    \end{align*}
At subsequent orders the leading order operator $(1 - 6 \phi_0^\pm + {6\phi_0^\pm}^2)$ pairs with the highest order $\phi^\pm_k$.
The forcing terms contain lower order factors.
As the first inhomogeneity is zero all remaining forcing terms, and therefore all outer phase-field expansions $\phi^\pm_{k\geq 1}$, are zero.

\emph{Fluid problem} --- The remaining fluid equations are linear and homogeneous,     
	\begin{align*}
    \partial_t T^+_1 + u^+_0\cdot\nabla T^+_1 + u^+_1\cdot\nabla T^+_0 - \kappa \lapof T^+_1 =0, \qquad
    \partial_t C^+_1 + u^+_0\cdot\nabla C^+_1 + u^+_1\cdot\nabla C^+_0 - \mu \lapof C^+_1 = 0,\\
    \partial_t u^+_1 + u^+_0\cdot\nabla u^+_1 + u^+_1\cdot\nabla u^+_0 - \nu \lapof u^+_1 + \nabla p^+_1 - B(T^+_1-N C^+_1)\hat{z}= 0, \qquad\qquad
    \nabla \cdot u^+_1 = 0.
    \end{align*}
The external boundary conditions are satisfied by the zeroth order outer solutions. 
The first-order corrections therefore have homogeneous external boundaries.
They are only perturbed at the melting interface.
We show this interfacial perturbation is zero at first order.
    
\emph{Solid problem} --- The solid equations are the same at first order,
	\begin{align*}
    \partial_t T^-_1 - \kappa \lapof T^-_1 &= 0, &
    \partial_t C^-_1 - \mu \lapof C^-_1 &= 0, &
	u^-_1 &= 0, &
	\nabla^2 p^-_1 &= B\partial_z (T^-_1 - N C^-_1).
    \end{align*}
    
\emph{Boundary problem} --- 
The differential operator of the first-order phase-filed equation has a two dimensional kernel spanned by $\dx\phi_0$ and ${6\phi_0(1-\phi_0)(\xi + \sinh \xi) + \sinh \xi}$.
The inhomogeneity must be orthogonal to the kernel, implying the Gibbs-Thomson condition,
    \begin{align*}
    \gamma (\dx^2 -(1-6\phi_0+6\phi^2_0))\phi_1 &= -\dx\phi_0(T_0 + mC_0 - \gamma \Kbar) &
    \implies &&
	(T_0^\pm + mC_0^\pm)|_{\sigma=0} - \gamma \Kbar &= 0.
	\end{align*}
Matching implies $\phi_1 = 0$.	
Integrating the temperature equation recovers energy conservation,
	\begin{align*}
    \kappa \dx^2 T_1 &= v_0 L \dx \phi_0,&
    \implies &&
	T_1 &= \frac{-v_0 L}{\kappa} \int_\xi^\infty \phi_0 \, d\eta + \partial_\sigma T_0^+(0) \xi + T_1^+(0), &
	[\kappa \partial_\sigma T^\pm_0]_{\sigma=0} + L v_0 &= 0.
	\end{align*}
Solving the homogeneous concentration equation and matching gives the solute conservation condition,
	\begin{align*}
	(\dx + \phi_0)(\mu \dx C_1 + v_0 C_0) &= 0, &
	\implies & &
	C_1 &= \frac{-v_0}{\mu} C^+_0(0) \xi + C^+_1(0), &
	\mu \partial_\sigma C^+_0|_{\sigma=0} + C^+|_{\sigma=0} v_0 &= 0.
	\end{align*}
The divergence condition when matched with the solid again implies zero normal velocity,
	\begin{align*}    
	\dx u_{\sigma 1} &= 0, & 
	\implies &&
	u_{\sigma 1} &= u^\pm_{\sigma 1}|_{\sigma=0} = 0.
	\end{align*}
The tangential velocity is proportional to the physical solution $\mathcal{U}(\xi)$.
We choose $\beta$ \cite{HesterImprovingConvergenceVolume2019} to ensure linear behaviour in the fluid,
	\begin{align*}
	\br{\dx^2 - \frac{\phi_0}{\beta^2}} u_{\bot 1} &= 0 &
	\implies & & 
	u_{\bot 1}(\xi) &= \partial_\sigma u^+_{\bot0}|_{\sigma=0} \mathcal{U}(\xi), &
	u_{\bot 1}(\xi\to\infty) &\sim \partial_\sigma u^+_{\bot0}|_{\sigma=0} \xi, &
	u^+_{\bot 1}|_{\sigma=0} = 0.
	\end{align*}
We finally use the normal momentum equation to show the pressure is constant at leading order
	\begin{align*}
	\dx p_0 + \nu\frac{\phi_0}{\beta^2}u_{\sigma 1} &= 0 &
	\implies &&
	p_0 &= p^\pm_0(0).
	\end{align*}
The first-order asymptotics have reproduced the Gibbs-Thomson condition with zero interface kinetics and the solute and energy conservation boundary conditions.
Hence the phase-field equations will tend to the sharp interface equations in the limit.
Calibrating $\beta$ ensures the first-order outer fluid velocity satisfies homogeneous boundary conditions $u^+_{\sigma 1}|_{\sigma=0} = u^+_{\bot 1} |_{\sigma =0} = 0$.
We now solve the second-order asymptotics to show that the mobility coefficient $\eps (5/6)(L/\kappa)$ ensures homogeneous boundary conditions for the first-order outer temperature and concentration, and that the first-order interfacial velocity error is zero --- allowing second-order convergence.

\subsection{Second order}
\label{sec:second}
\emph{Fluid problem} --- At second order the fluid equations are sourced by the first-order errors,
    \begin{align*}
    \partial_t T^+_2 + u^+_0\cdot\nabla T^+_2 + u^+_2\cdot\nabla T^+_0 - \kappa \lapof T^+_2 &= - u^+_1\cdot\nabla T^+_1,\\
    \partial_t C^+_2 + u^+_0\cdot\nabla C^+_2 + u^+_2\cdot\nabla C^+_0 - \mu \lapof C^+_2 &= - u^+_1\cdot\nabla C^+_1,\\
    \partial_t u^+_2 + u^+_0\cdot\nabla u^+_2 + u^+_2\cdot\nabla u^+_0 - \nu \lapof u^+_2 + \nabla p^+_2 - B(T^+_2-N C^+_2)\hat{z}&= -u^+_1\cdot\nabla u^+_1,\\
    \nabla \cdot u^+_2 &= 0.
    \end{align*}    
\emph{Solid problem} --- The solid velocity is now non-zero from interior pressure and forces,
	\begin{align*}
    \partial_t T^-_2 &= \kappa \lapof T^-_2, &
    \partial_t C^-_2 &= \mu \lapof C^-_2, &
    \nabla^2 p^-_2 &= B\partial_z (T^-_2 - N C^-_2), &
    \frac{\nu}{\beta^2} u^-_2 &=  B(T^-_0 - N C^-_0)\hat{z} -\grad p^-_0.
    \end{align*}
\emph{Boundary problem} --- The phase-field equation now has a more complex inhomogeneous term.
	\begin{align*}
	\gamma(\dx^2 - (1-6 \phi_0 + 6\phi^2))\phi_2 &=  -\dx \phi_0\br{ \frac{5}{6}\frac{L}{\kappa} v_0 + (T_1+mC_1) - \xi \gamma\overline{K^2}} .
	\end{align*}
We again apply a solvability condition.
The odd terms drop out, and integration shows ${\int_{-\infty}^\infty -\dx \phi_0 \phi_0(1-\phi_0) \, d\xi = \frac{1}{6}}$ and
${\int_{-\infty}^\infty -\dx \phi_0 \phi_0(1-\phi_0) \int_\xi^\infty \phi_0\, d\eta\, d\xi = \frac{5}{36}}$.
The interfacial velocity term thus vanishes, giving a constraint between first-order concentration and temperature.
	\begin{align*}
	\int_{-\infty}^{\infty} \dx \phi_0^2 \br{T_1^\pm(0) + m C_1^\pm(0) + \frac{L v_0}{\kappa} \br{\frac{5}{6} - \int_\xi^\infty\phi_0 \, d\eta}}d\xi &= 0 &
	\implies &&
	(T_1^\pm + m C_1^\pm)|_{\sigma=0} &= 0.
	\end{align*}
An explicit formula for $\phi_2$ (using variation of parameters) is non-trivial but it decays exponentially to zero in either direction.
The temperature equation is then integrated.
If initially $T^\pm_1|_{\sigma=0}=0$, then matching prevents linear asymptotic behaviour of $T_2$, implying $v_1 = 0$,
	\begin{align*}
	\kappa \dx^2 T_2 &= L v_1 \dx \phi_0 + \partial_t T_0 - \kappa \sdiv \sgrad T_0 +(\kappa \Kbar- v_0) \dx T_1,\\
	\kappa T_2 &= 
	(\partial_t - \kappa \sdiv \sgrad + (\kappa \Kbar - v_0)\ds)T_0^+|_{\sigma=0} \frac{\xi^2}{2}
	+(\kappa \Kbar - v_0)\frac{v_0 L}{\kappa} \int_{\xi}^{\infty}\int_{\eta}^\infty \phi_0 \, d\zeta \, d \eta + T^\pm_2|_{\sigma=0}.
	\end{align*}
The concentration equation is similarly integrated. 
The inhomogeneity is constant in $\xi$, and explicitly solvable.
Using previous solutions and $C^\pm_1|_{\sigma=0} = T^\pm_1|_{\sigma=0} = v_1 = 0$, we find $C_2$,
	\begin{align*}
	(\dx + \phi_0)(\mu \dx C_2 + v_0 C_1 + v_1 C_0) &= \partial_t C_0 + \mu \Kbar \dx C_1 - \mu \sdiv \sgrad C_0,\\
	\mu C_2 = \frac{v_0^2}{\mu^2}C_0^+|_{\sigma=0}\frac{\xi^2}{2} + (\partial_t + \Kbar v_0 &- \mu\sgrad \cdot \sgrad )\,C^\pm_0|_{\sigma=0} \int_0^\xi \frac{-\log\phi}{1-\phi} d\eta + C_2^\pm|_{\sigma=0}.
	\end{align*}
The divergence equation implies the second-order normal velocity is now no longer zero, 
	\begin{align*}
	\dx u_{\sigma2} &= - \sdiv u_{\bot1}, &
	\implies &&
	u_{\sigma 2} &= u_{\sigma 2}(-\infty) - \sdiv{(\ds u_0^+(0))} \int^\xi_{-\infty} \U \, d\zeta.
	\end{align*}
The second-order tangential velocity can then be solved using variation of parameters
    \begin{align*}
	\nu\br{\dx^2 - \frac{\phi_0}{\beta^2}}u_{\bot 2} &= \sgrad p_0 + (\nu \Kbar - v_0) \dx u_{\bot 1} + B(T_0 - NC_0)\hat{z}_\bot = \mathcal{R} ,\\
    u_{\bot 2} &= (\mathcal{Q} + c)\, \U, \quad \text{where} \quad \mathcal{Q} \equiv \int^\xi_0 -\frac{\int_{-\infty}^\eta \mathcal{R}{\U} d\zeta}{\U^2}d\eta, 
	\end{align*}
where $c$ is a constant of integration that cancels the linear part of the solution into the fluid.
The normal momentum equation can then be integrated to determine $p_1$.
Matching requires no constant term in the limiting behaviour of the pressure into the fluid,
	\begin{align*}
	\dx p_1 &= \nu \sdiv (\ds u_{\bot0}^+|_{\sigma=0})\br{\frac{\phi_0}{\beta^2}\br{\int_{-\infty}^{\xi}\U\, d\eta - \U'(\xi)}} -  \frac{\nu\phi_0}{\beta^2}u_{\sigma 2}^-|_{\sigma=0} - B (T_0^+ - N C_0^+)|_{\sigma=0}\hat{z}_\sigma ,\\
	p_1 &= \nu \sdiv (\ds u_{\bot0}^+|_{\sigma=0})\U(\xi) - B  (T_0^+ - N C_0^+)|_{\sigma=0} \hat{z}_\sigma \xi,\\
	 & \quad +\frac{\nu}{\beta^2}\br{ \sdiv (\ds u_{\bot0}^+|_{\sigma=0})\br{\int_\xi^\infty \phi_0 \int_{-\infty}^\zeta \U(\eta) \, d\eta \, d\zeta } + u_{\sigma 2}^-|_{\sigma=0}\int_\xi^\infty\phi_0\, d \eta}.
	\end{align*}

The second-order asymptotic analysis shows that calibrating the mobility and damping parameters ensures homogeneous boundary conditions of the first-order outer solutions at the interface.
Combined with homogeneous external boundary conditions and homogeneous linear evolution equations at first order, 
this implies that if the outer solutions are initialised correct to $\O(\eps^2)$, 
then the fields and interfacial velocity will evolve accurate to $\O(\eps^2)$ over time.
In reality the chaotic nature of many fluid dynamics problems prevents convergence beyond the Lyapunov timescale of the flow, but at each point in time the system behaves correctly to within second-order accuracy.

\section{Numerical validation of the model}
\label{sec:experiments}
We now validate the asymptotic arguments of \cref{sec:asymptotics} in two benchmark problems.
In each problem we calculate a numerical reference solution corresponding to the ``sharp interface'' equations of \cref{sec:sharp}.
We then show the optimal phase-field equations of \cref{sec:smooth} achieve convergence of $\O(\eps^2)$ to these reference solutions.
Both the reference and phase-field problems are simulated using the flexible and efficient spectral code Dedalus \cite{BurnsDedalusFlexibleFramework2020}\footnote{The full simulation code, saved data, and analysis scripts are freely available at \href{https://github.com/ericwhester/phase-field-code}{github.com/ericwhester/phase-field-code}.}.
This allows us to probe equation-level model error in the absence of numerical discretisation errors.

\subsection{Melting and dissolution at a stagnation point}
\label{sec:salty-stag}
\begin{figure}[t]
        \centering
        \includegraphics[width=\linewidth]{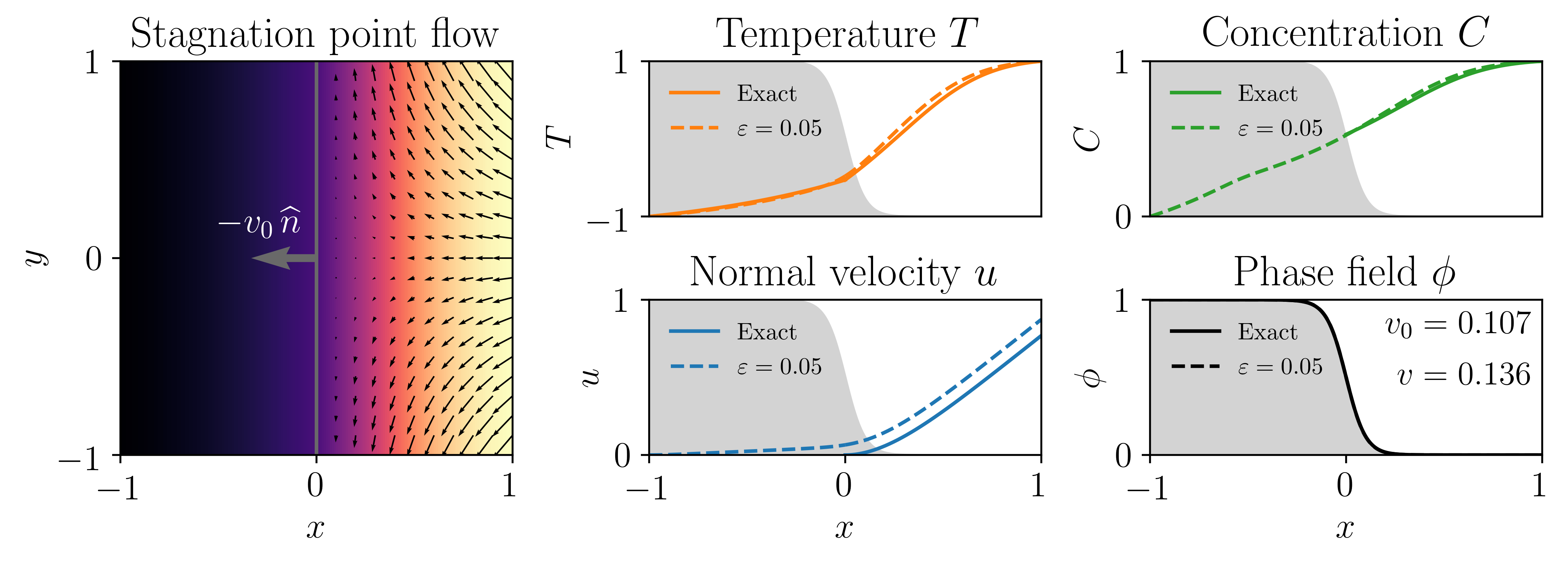}
        \caption{
        On the left we illustrate stagnation point flow toward a melting interface. 
        The fluid velocity decreases to zero at the interface (black arrows $x>0$), and the vertical velocity is proportional to $y$.
        The interface velocity (grey arrow at $x=0$) uniformly recedes to the left.
        The temperature (in colour) decreases toward the left, and is invariant in $y$ (as are all quantities excepting the vertical fluid velocity).
        The concentration (not shown) decreases similarly.
        We also plot reference and phase-field solutions for temperature $T$, solute concentration $C$, normal velocity $u$, and phase field $\phi$ as a function of $x$ in the second and third columns.
        Only the temperature exists inside the solid in the reference solution.
        We negate the normal fluid velocity $u$ for clarity.}
        \label{fig:melt-salt-flow-diagram}
    \end{figure}

The first benchmark problem examines warm liquid with dissolved solute flowing toward a melting interface at a stagnation point, (similar to section 5 of \cite{LeBarsInterfacialconditionspure2006} and 5.6 of \cite{HesterImprovingConvergenceVolume2019}).
We compare the sharp interface and phase-field approximations to this problem, as illustrated in \cref{fig:melt-salt-flow-diagram}.
The symmetries of the system allow us to significantly simplify \cref{eq:sharp-pdes},
revealing a steady travelling wave similarity solution for the moving boundary and phase-field formulations.
We do this by transforming to a frame moving leftward at the steady interface melting speed $-v$.
We solve the system as a nonlinear boundary value problem in Dedalus.

\emph{Sharp interface model} --- We solve a diffusion equation for the solid temperature $T^-$, advection-diffusion equations for the liquid temperature $T^+$ and solute concentration $C$, and a nonlinear third order equation for the horizontal fluid velocity $u$, 
    \begin{align} 
	\kappa\partial_x^2 T^+ &= (u-v)\partial_x T^+, &
	\mu\partial_x^2 C &= (u-v)\partial_x C,\nonumber \\
	\kappa\partial_x^2 T^- &= -v\partial_x T^-, &
    \nu \partial_x^3\notvc{u}&= 1 + (u-v)\partial_x^2\notvc{u}- (\partial_x u)^2.
    \end{align}
Note that though we solve in the moving frame, we do not apply a Galilean boost to the fluid velocity.
These equations must also satisfy energy conservation, solute conservation, temperature continuity and Gibbs-Thomson relations at the liquid-solid interface, and Dirichlet conditions at $ x \pm 1$.
We can then solve for each variable and the unknown melting speed $v$.
    \begin{align}
    T^-(-1) &= -1, 
    &
    T^+(0) &= T^-(0),
    &
    [\partial_x T(0)]^+_- &=- \frac{L v}{\kappa},
    &
	u(0) &= 0,
	&
	\partial_x u(0) &= 0,\nonumber\\
    T^+(1) &= 1,
    &
	T^\pm(0) &=- m C(0),
	&
	\partial_x C(0) &= -\frac{C(0)v}{\mu},
	&
	C(1) &= 0,
	&
	\partial_x u(1) &= -1, 
    \end{align}

\emph{Phase-field model} --- The phase field instead implicitly models the interfacial boundary conditions through various equation terms, which reduce \cref{eq:phase-pdes} to,
    \begin{align}
	\nu\partial_x^2 T &= ((1-\phi)u-v)\partial_x T + Lv\partial_x \phi,\nonumber\\
	\mu\partial_x^2 C &= (u-v)\partial_x C -\partial_x \log(1-\phi + \delta)\left(\mu\partial_x C + v C\right), \nonumber\\
    \nu \partial_x^3\notvc{u}&= 1 + (u-v)\partial_x^2\notvc{u}- (\partial_x u)^2 + \frac{\nu}{(\beta\eps)^2} \phi \partial_x u,\nonumber\\
    \gamma \partial_x^2 \phi &= -\eps\frac{5}{6}\frac{L}{\kappa} v \partial_x \phi + \frac{\gamma}{\eps^2} \phi(1-\phi)(1-2\phi) + \frac{1}{\eps}\phi(1-\phi)(T + mC).
    \end{align}
We now only require the Dirichlet outer boundary conditions of before to solve the problem
    \begin{align}
	T(-1) &= -D, & 
	C(-1) &= 0,	&
	\phi(-1) &= 1, &
	u(-1) &= 0,	&
	\partial_x u(-1) &= 0, \nonumber\\
	T(1) &= 1, &
	C(1) &= 1, &
	\phi(1) &= 0, &
	\phi(0) &= 1/2, &
	\partial_x u(1) &= -1.
	\end{align}
Requiring $\phi(0)=1/2$ is necessary to fix the problem in the moving frame to determine $v$.
    \begin{figure}[ht]
        \centering
        \includegraphics[width=.6\linewidth]{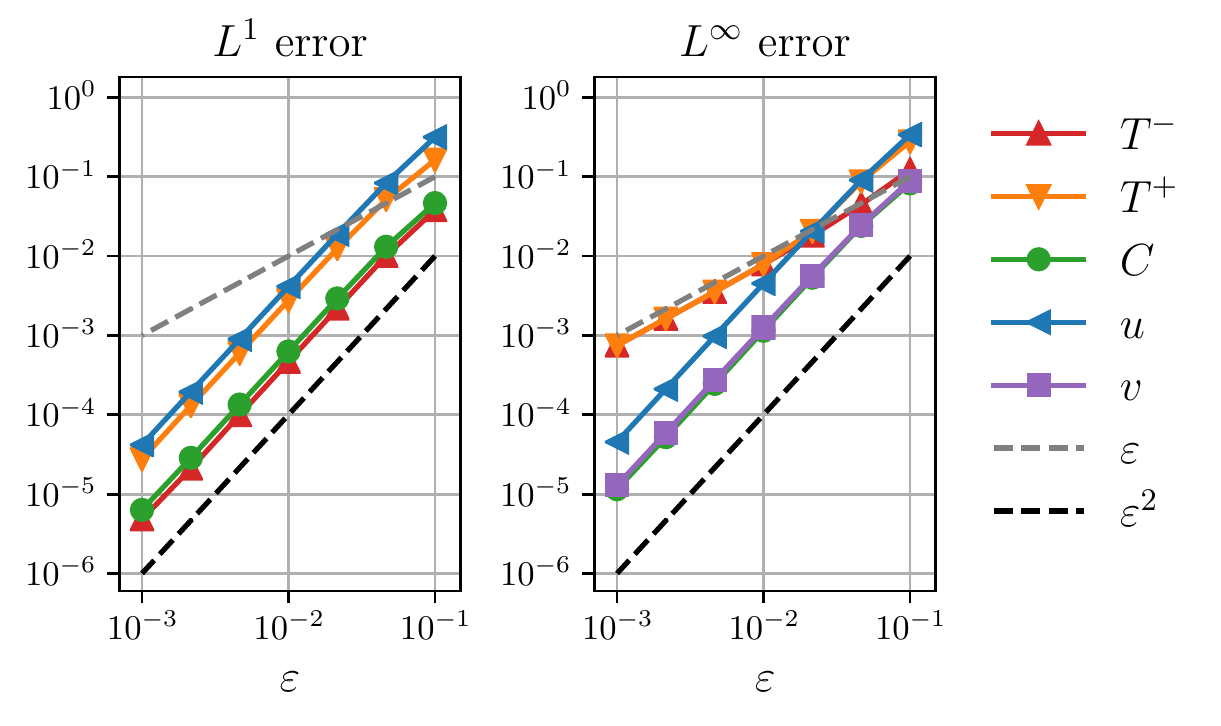}
        \caption{Plot of convergence in $L^1$ and $L^\infty$ error norm of the velocity $u$, liquid temperature $T^+$, solid temperature $T^-$, solute concentration $C$ and interface melting speed $v$ as a function of $\eps$. Each norm is calculated within the appropriate domain of the reference variable. Clear $\O(\eps^2)$ convergence is observed in $L^1$ norm using optimal parameters. $\O(\eps)$ convergence in $L^\infty$ norm occurs for the temperature because of the non-differentiability of the reference solution.}
        \label{fig:melt-salt-flow-convergence}

    \end{figure}
	
\emph{Results} --- We solve each problem as a nonlinear boundary value problem in Dedalus.
We discretise each variable on the solid ($-1<x<0$) and fluid ($0<x<1$) domains using Chebyshev polynomials.
Newton-Kantorovich iteration then converges on a solution with a tolerance of $10^{-12}$ (see \cite{BoydChebyshevFourierSpectral2001} appendix C).
We reproduce example phase field and reference snapshots in \cref{fig:melt-salt-flow-diagram} for $\eps = 0.05$, $\kappa = \mu = \nu = 1/10$, and $D = m = L = \gamma = 1$, using 64 grid points for each subdomain.
We set $\delta = 2\times 10^{-5}$ to regularise the solute equation within the solid.
The disagreement (though small) is visible by eye for the temperature, concentration, and normal velocity.

We then perform a quantitative analysis of convergence in \cref{fig:melt-salt-flow-convergence}.
We test seven logarithmically spaced values of $\eps$ from $10^{-1}$ to $10^{-3}$, for the previous control parameters, and using 256 grid points in each subdomain.
We quantify the difference between the reference and phase-field solutions with the $L^1$ and $L^\infty$ error norms of each variable $(u, T^+, T^-, C, v)$ as a function of $\eps$.
We find clear $\O(\eps^2)$ convergence in the $L^1$ error norm of each field variable, as well as for the $L^\infty$ error norm of the $u$, $C$ and $v$ variables.
The $L^1$ convergence demonstrates the quantitative accuracy of the phase-field model.
The reason for the apparently restricted $\O(\eps)$ $L^\infty$ convergence of the temperature fields is the jump in temperature gradient of the reference solution at the interface.
The smooth phase-field model cannot follow this kink leading to an $\O(\eps)$ disagreement that is localised to the boundary.
Tangential velocities also suffer from this reduced continuity in general.
%

\subsection{Double diffusive melting}
\label{sec:salty-boussinesq}
In our second benchmark we examine buoyancy driven flow of warm liquid with dissolved solute underneath a melting solid layer.
This problem develops nontrivial geometries from the evolution of the flow.
To simulate the reference formulation in Dedalus, we use an evolving coordinate system that maps the fluid and solid regions to a stationary rectangular domain.
This transformation allows an efficient spectral discretisation using Dedalus, and is used in similar spectral solvers \cite{SubichSimulationNavierStokes2013}.
We repeat this remapping for the phase-field formulation as it concentrates resolution near the $\phi=1/2$ level set.
This allows us to efficiently and accurately simulate much smaller $\eps$.
By comparing the phase-field simulation to the reference problem as we decrease $\eps$, we demonstrate second-order accuracy of the model.

\emph{Sharp differential equations} --- The full domain exists between $0 < z < 2$ and $0 < x < 4$.
It is partitioned by the interface at height $z = h(t,x)$.
Above the interface we solve the heat equation for the solid temperature field $T^-$ with diffusivity $\kappa$.
Below the interface we solve incompressible Boussinesq hydrodynamics (using a first-order formulation in terms of velocity $u$, vorticity $q$, and augmented pressure $p = P + \tfrac{1}{2}|u|^2$), with advection and diffusion of temperature $T^+$ and solute concentration $C$,
	\begin{align}
	\partial_t T^- - \kappa \nabla^2 T^- &= 0, &
	\nabla \cdot\notvc{u}&= 0,\nonumber\\
	\partial_t T^+ - \kappa \nabla^2 T^+ &= - u\cdot \nabla T^+, &
	q - \nabla \times\notvc{u}&= 0,\nonumber\\
	\partial_t C - \mu \nabla^2 C &= - u\cdot \nabla C, &
	\partial_t\notvc{u}+ \nabla p - \nu \nabla \times q &= q \times\notvc{u}+ (T^+ + N C)\,  \ehat_z,
	\end{align}
where $\nu,\kappa,$ and $\mu$ are the momentum, temperature, and solute diffusivity.
This first-order reformulation is required in the Dedalus solver, and has the added benefit of simplifying the mathematical details of the coordinate transformation.
We set the buoyancy parameter $B = 1$.

\emph{Boundary conditions} --- 
At the top we specify conservative temperature boundary conditions.
At the bottom, we specify no-slip boundaries with no-flux temperature and solute conditions.
The zeroth mode for the vertical velocity is replaced by a choice of pressure gauge,
	\begin{align*}
	\partial_z T^-(t,x,2) &= 0, &
	\partial_z T^+(t,x,0) &= 0, &
	\partial_z C(t,x,0) &= 0, &
	u(t,x,0) &= 0, &
	p(t,x,0) &= 0.
	\end{align*}
At the interface $z = h(t,x)$, we have a Gibbs-Thomson boundary condition, matching temperature boundary conditions, energy and solute conservation boundary conditions, and zero velocity boundary conditions required for mass conservation,
	\begin{align}
	T^+ + m C &= -\gamma \nabla \cdot \nhat, &
		\nhat \cdot \nabla T^+ - \nhat \cdot \nabla T^- + \frac{L}{\kappa} \nhat \cdot \partial_t h \, \ehat_z &= 0, &
		u_x &= 0, \nonumber\\
	T^+ - T^- &= 0,
		 &
	\nhat \cdot \nabla C + \frac{C}{\mu} \nhat \cdot \partial_t h \, \ehat_z &= 0, &
	u_z &= 0. 
	\end{align}

\emph{Initial conditions} --- We initialise the problem with $h(0,x) = 1$ and zero velocity and pressure, and a decreasing concentration profile with height.
We apply a large perturbation to the linearly decreasing temperature field to initiate convection,
	\begin{align*}
	T(0,x,z) &= 1 - z + \exp(5^2((x-2)^2 + (z-0.5)^2)),\\
	C(0,x,z) &= 0.05 + (1-0.05)\frac{1}{2}\br{1 - \tanh(10(z-0.5))}.
	\end{align*}

\emph{Phase-field equations} --- The phase-field equations are a similar reformulation of \cref{eq:phase-pdes}.
	\begin{align}
    \partial_t T  - \kappa \lapof T - L \partial_t \phi &= - u\cdot\nabla T,\nonumber &
	\partial_t C  - \mu \nabla^2 C &= - u\cdot \nabla C + \frac{\partial_t \phi C - \nabla C \cdot \nabla \phi}{1-\phi + \delta}\nonumber\\
	\nabla \cdot\notvc{u}&= 0,\nonumber &
	\eps\frac{5}{6}\frac{L}{\kappa}\partial_t\phi -\gamma \lapof \phi  & = - \frac{1}{\eps^2}\phi(1-\phi)(\gamma (1 - 2\phi) +\eps (T+mC)),\\
	q - \nabla \times\notvc{u}&= 0,\nonumber &
	\partial_t\notvc{u}+ \nabla p - \nu \nabla \times q &= q \times\notvc{u}+ (T^+ + N C)\,  \ehat_z - \frac{\nu}{(\beta\eps)^2} \phi \, u,\nonumber\\
	\end{align}
We use the optimal damping prescription $\beta = 1.51044385$ \cite{HesterImprovingConvergenceVolume2019} and choose $\delta = 10^{-4}$. 
We specify the same insulating no-slip boundary conditions at $z = 0$ and $z = 2$.
The boundary conditions at $z = h(t,x)$ require continuity of each field and its derivatives.
We initialise using the same initial conditions as for the remapped simulation, plus the initial conditions for the phase field
	\begin{align*}
	\phi(0,x,z) = \frac{1}{2}\br{1 + \tanh\br{\frac{1}{2\eps}}(z - 1)}.
	\end{align*}
	
\begin{figure}
	\centering  
    \includegraphics[width=.6\linewidth]{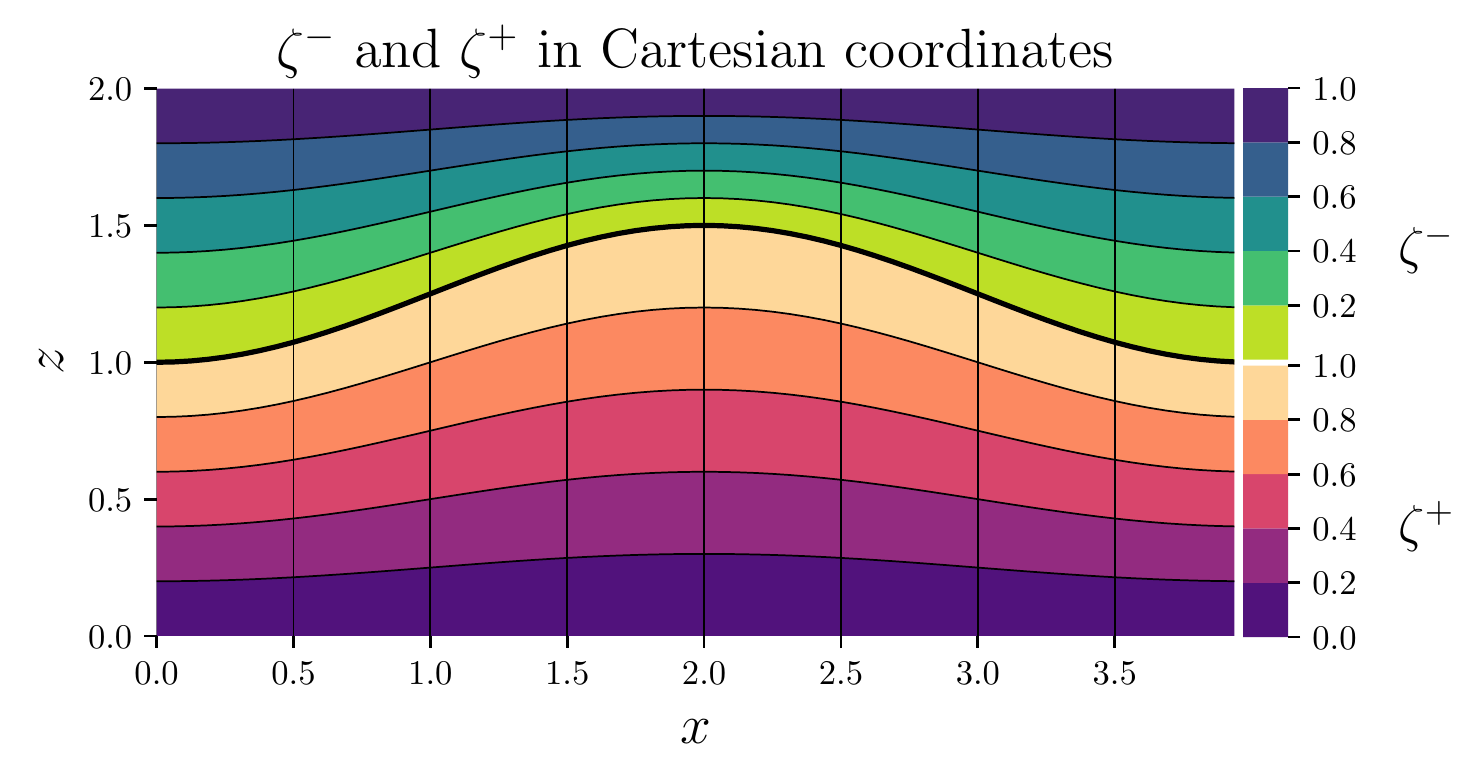}
    \caption{Contour plot of $\xi$ (vertical) and $\zeta$ coordinates (horizontal) for $h(x) = \frac{3}{4}+ \frac{1}{5}\cos(\pi x)$.}
    \label{fig:zeta-cartesian}
\end{figure}

\emph{Remapped coordinates} --- To solve the melting problem we remap our evolving domain in Cartesian space to a fixed rectangular domain with the new coordinates $\tau, \xi,$ and $\zeta^\pm$,
	\begin{align*}
	\tau(t,x,z) &= t, &
	\xi(t,x,z) &= x, &
	\zeta^+(t,x,z) &= \frac{z}{h(t,x)}, &
	\zeta^-(t,x,z) &= \frac{z-h}{2-h(t,x)},
	\end{align*}
for which we plot equally spaced level set contours in \cref{fig:zeta-cartesian}.
The differential geometry of the new coordinates is presented in \cref{app:remapped-coordinates}.
We solve the phase field equations on the remapped domains of the reference problem to concentrate resolution near the phase-field interface and speed up comparison of quantities between simulations.

\emph{Model and numerical parameters} --- We simulate these equations using model parameter values from  \cref{tab:salty-boussinesq-melting-params}.
We simulate the reference equations using 64 Chebyshev polynomials in the vertical direction and 128 Fourier modes in the periodic horizontal direction.
After determining the reference solution, we discretise the phase-field equations on the same evolving domain of the reference simulation for several values of $\eps$.
To interpolate the reference geometries into the phase-field simulations between the saved time and grid points we use third order interpolating splines.
The phase-field simulations discretise the vertical $\zeta$ basis using three compound Chebyshev bases $[0,10\,\eps]\cup[10\,\eps,1-10\,\eps]\cup[1-10\,\eps,1]$, with resolutions of 32, 64, and 32 modes respectively.
This greatly reduces the simulation cost.
We use a time step size of $\Delta t = 10^{-3},5\times 10^{-4},2.5\times 10^{-4},2\times 10^{-4},5 \times 10^{-5}$ for decreasing choice of $\eps$, and integrate in time using a second-order multistep semi-implicit backwards difference formula (SBDF2).
	\begin{table}[hbt]
	\centering
    \begin{tabular}{cccccccc}
    $\nu$ & $\kappa$ & $\mu$ & $\gamma$ & $L$ & $m$ & $N$ & $\eps$\\
    \hline 
    $10^{-2}$ & $10^{-2}$ & $10^{-2}$ & $10^{-2}$ & 1 & 0.2 & 0 & $\sqrt{5}\cdot 10^{-3},10^{-2.5},\sqrt{2}\cdot 10^{-2.5},\sqrt{5}\cdot 10^{-2.5},10^{-2}$
    \end{tabular}
    \caption{Model parameters used in the second benchmark problem.}
    \label{tab:salty-boussinesq-melting-params}
	\end{table}

\begin{figure}
        \centering
        \includegraphics[width=\linewidth]{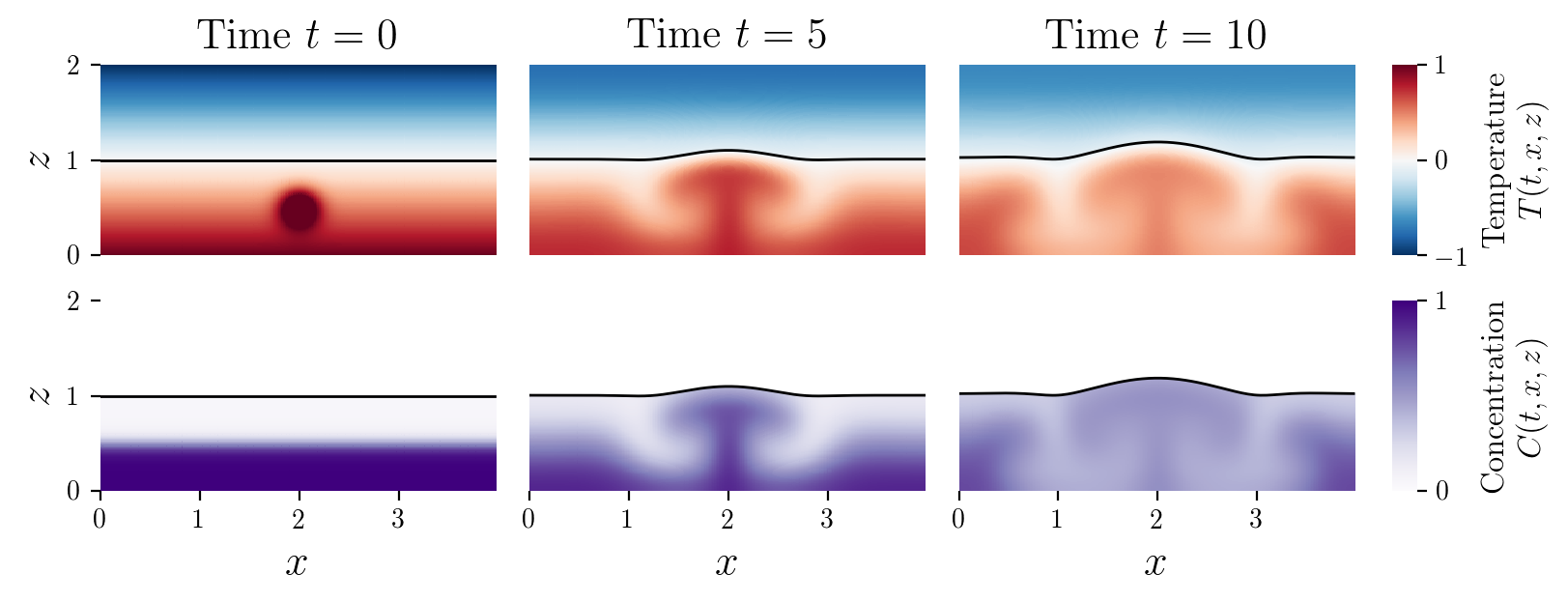}
        \caption{Time series of temperature and concentration fields in reference simulation.}
        \label{fig:temp-salt-time-series}
		\vspace{1em}
        \includegraphics[width=\linewidth]{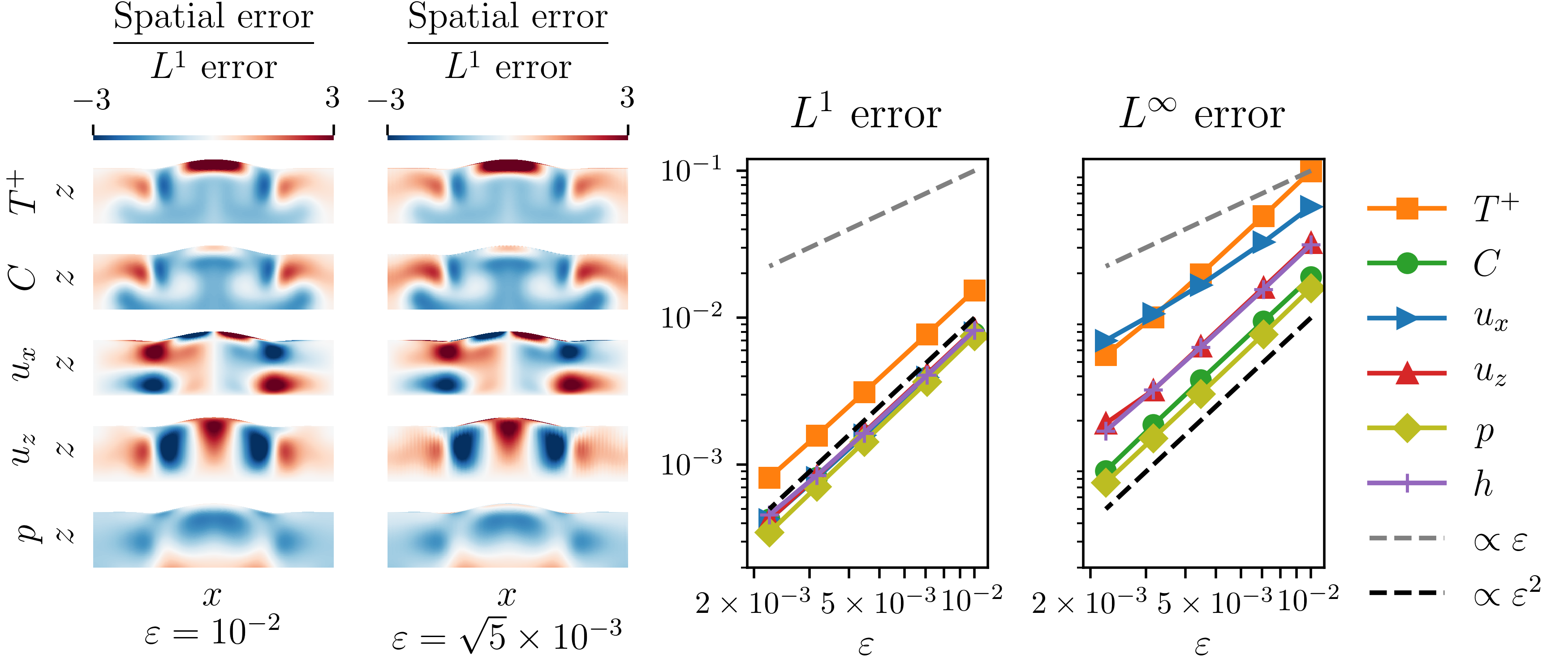}
        \caption{Column 1 and 2 plot normalised spatial errors of the liquid temperature $T^+$, concentration $C$, horizontal and vertical Cartesian velocity $u_x, u_z$, and pressure $p$ at time $t = 10$. 
        Column 3 and 4 plot the $L^1$ and $L^\infty$ error norms for the fluid temperature $T^+$, concentration $C$, vertical $u_x$ and horizontal velocity $u_z$, pressure $p$, and interface height $h$ as a function of $\eps$ at time $t = 10$.}
        \label{fig:salty-boussinesq-melting-errors}        
    \end{figure}

\emph{Results} --- 
A time series of the temperature and concentration fields of the reference solution is given in \cref{fig:temp-salt-time-series}, which illustrates a rising buoyant plume from the initial temperature anomaly in the fluid. The warm solute-laden liquid melts the interface in the middle more rapidly than the ambient liquid at the sides, causing a trough to develop.

In \cref{fig:salty-boussinesq-melting-errors} we plot several error metrics of the phase-field simulations.
In the first two columns we plot the spatial error normalised by the $L^1$ error for the liquid temperature $T^+$, solute concentration $C$, Cartesian velocity components $u_x$ and $u_z$, and true pressure $P = p - \tfrac{1}{2}|u|^2$.
We plot these normalised spatial errors for the smoothest ($\eps = 10^{-2}$) and sharpest ($\eps = \sqrt{5}\times 10^{-3}$) phase-field simulations at the final time $t = 10$.
These plots reveal a consistent spatial error profile between simulations.
To understand the amplitude of the spatial error, we plot the $L^1$ and $L^\infty$ error norms of each variable as a function of $\eps$ in the third and fourth columns of \cref{fig:salty-boussinesq-melting-errors}.
We see clear second-order convergence of all variables in $L^1$ norm.
We note that the structure of the boundary layer of the tangential velocity and temperature affects $L^\infty$ error norm.
The phase-field model causes a kink in the temperature and tangential velocity near the interface, which leads to an $O(\eps)$ error of these variables in the interfacial region.
However, this error is localised to the boundary, and does not propagate outward.
(This trend is difficult to notice in the temperature plot as the error within the fluid still dominates the $\O(\eps)$ boundary error for the moderate choices of $\eps$ chosen.)
We therefore achieve second-order convergence in the fluid and solid regions due to our optimal calibration of the phase-field model parameters.

\section{Conclusions}
\label{sec:conclusion}
In this paper we provide a general framework for analysing the convergence of phase-field models.
We use this procedure to develop a second-order phase-field model of melting and dissolution in multi-component flows.
This is a concrete advancement that showcases second-order accurate approximations of many common boundary conditions; no-slip Dirichlet boundaries, Neumann boundaries, Robin boundaries, and Stefan boundaries.
We also verify these prescriptions in two thorough benchmark problems with accurate reference solutions.
By developing a framework to validate this model, we now possess the machinery requied to create second-order accurate extensions to more general thermodynamic properties.
We can also consider yet higher order analyses of this and other models.
An automated approach of Richardson sequence extrapolation could also be considered to generate higher order accurate models, as was done by the authors for the volume-penalty method \cite{HesterImprovingConvergenceVolume2019}.
To be clear, phase-field models are not necessarily the most appropriate choice for any problem.
Remapping was also shown to be an effective strategy for sufficiently simple geometries in \cref{sec:salty-boussinesq}.
However our second-order phase-field model is simple to implement, is more accurate than standard phase-field models, and is applicable in much more challenging geometries than remapping approaches.

\appendix
\section{Signed-distance coordinates}
\label{app:sdf}
The signed distance $\sigma$ of the point $x$ is the minimum distance to the surface.
Labelling surface points $p$ and unit normals $\nhat$ with orthogonal surface coordinates $s$, we have    \begin{align*}
    x = p(s) + \sigma\, \nhat(s).
    \end{align*}
This gives a tangent vector basis $t_i$, dual basis $\nabla s_i$, and orthonormal tangent basis $\that_i$.
    \begin{align*}
    {t}_i &= \frac{\partial {p}}{\partial s_i}, &
    \that_i &= \frac{t_i}{|t_i|}, &
    \nabla s_i &= \frac{\that_i}{|t_i|}.
    \end{align*}
These give the surface gradient $\sgrad$, area measure $dA = |t_1||t_2|ds_1ds_2$, and surface divergence,
    \begin{align*}
    \sgrad &= \nabla s_1  \frac{\partial}{\partial s_1} + \nabla s_2  \frac{\partial}{\partial s_2} = \that_1 \nabla_1 + \that_2 \nabla_2, &
    \sdiv u_\bot &= \frac{1}{|t_1||t_2|}\br{\pd{}{s_1}(|t_2| u_1) + \pd{}{s_2}(|t_1| u_2)}.
    \end{align*}
The normal $\nhat$ is everywhere equal to the gradient of the signed distance,
and is therefore its gradient is symmetric and diagonalisable.
The eigenvectors are the principal directions of curvature (which align with orthogonal surface coordinates), 
and the eigenvalues are the principal curvatures $\kappa_i$,
    \begin{align*}
    \nhat &= \nabla \sigma, &
    \nabla \nhat &= - \kappa_1 \that_1 \that_1 - \kappa_2 \that_2 \that_2 = - K.
    \end{align*}
We express the gradient using the orthonormal frame $(\nhat, \that_1, \that_2)$ and the scale tensor $J$
    \begin{align*}
    \nabla &= \nhat \ds + J^{-1} \cdot \nabla_\bot, \quad \text{where} \quad     J = I - \sigma K.
    \end{align*}
We now give the remaining geometric quantities needed for all vector calculus operations. 
The determinant $|J|$ relates mean curvature $\Kbar$, Gaussian curvature $\Kabs$, and volume measure $dV$, 
    \begin{align*}
    \Kbar &= \kappa_1 + \kappa_2, &
    \Kabs &= \kappa_1 \kappa_2, &
    |J| &= 1 - \sigma \Kbar + \sigma^2 \Kabs, &
    dV = |J| \, d\sigma \, dA.
    \end{align*}
We define the adjugate tensors $\Jhat$ and $\Khat$,     \begin{align*}
    \Khat &= |K| K^{-1} = \kappa_2 \that_1 \that_1 + \kappa_1 \that_2 \that_2,&
	\Jhat &= |J|J^{-1} = I - \sigma \Khat.
      \end{align*}
We also define the cross product with the unit normal, which admits simple identities,
    \begin{align*}
    \nabla^\bot &= \nhat \times \nabla_\bot, & 
	    \nabla^\bot \cdot \nabla_\bot &= \nabla_\bot \cdot \nabla^\bot = 0, & 
	    \nhat \times (J u_\bot) &= \Jhat u^\bot, \\
	u^\bot &= \nhat \times u_\bot, &
    	u^\bot \cdot u_\bot &= u_\bot \cdot u^\bot = 0, &
    	\nhat \times u^\bot &= -u_\bot,
    \end{align*}
We next find the gradient of the basis vectors, and define Ricci rotation coefficients,
    \begin{align*}
    \sgrad \nhat &= - \kappa_1 \that_1\that_1 -\kappa_2 \that_2 \that_2, & 
    \mathcal{R}_i^{jk} &= \that_j \cdot (\nabla \that_i) \cdot \that_k,
    	&
	 \mathcal{R}^{12}_1 &= \that_1 \cdot (\nabla \that_1) \cdot \that_2 =  \omega_1,
    \\
    \sgrad \that_i &= \kappa_i \that_i \nhat + \mathcal{R}^{j k}_i \that_j \that_k,
    &
    \mathcal{R}^{jk}_i &= - \mathcal{R}^{ji}_k, &
    \mathcal{R}^{21}_2 &= \that_2 \cdot (\nabla \that_2) \cdot \that_1 = -\omega_2.
    \end{align*}
These relations allow us to calculate all relevant vector calculus operators,
    \begin{align*}
      \divof{u} &= \frac{\ds (|J| u_\sigma)}{|J|} + \frac{\nabla_\bot \cdot (\Jhat u_\bot)}{|J|}, \\
    \lapof f &= \frac{\ds(|J|\ds f)}{|J|} + \frac{\sdiv(\Jhat J^{-1} \sgrad f)}{|J|},\\
    \nabla \times\notvc{u}&= -\nhat \frac{\sdiv(\Jhat u^\bot)}{|J|} + \Jhat^{-1} (\ds (\Jhat u^\bot) - \grads u_\sigma),\\
    -\nabla \times \nabla\notvc{u}&= 
    \frac{\nhat}{|J|} \br{ - \sdiv(\Jhat J^{-1}(\ds(J u_\bot) - \sgrad u_\sigma )}\\
    &\quad+ \Jhat^{-1} \ds\br{\Jhat J^{-1} (\ds(J u_\bot) - \sgrad u_\sigma)} + \Jhat^{-1} \grads\br{\frac{\divs(J u_\bot)}{|J|}},\\
    \grad\notvc{u}&= \nhat \,\nhat \,\ds u_\sigma + \nhat \ds u_\bot + J^{-1}(\sgrad u_\sigma+ K u_\bot)\,\nhat + J^{-1} (\sgrad u_\bot - K u_\sigma),\\
	u\cdot \nabla f &= u_\sigma \ds f + u_\bot \cdot J^{-1} \sgrad f,\\
	u \cdot \grad\notvc{u}&= \nhat \,\big(u_\sigma \ds u_\sigma+ u_\bot \cdot J^{-1} (\sgrad u_\sigma+ K u_\bot)\big) + u_\sigma \ds u_\bot + u_\bot \cdot J^{-1} (\sgrad u_\bot - K u_\sigma),\\
    \partial_t f &= \partial_\tau f - v \partial_\sigma f + \sigma \sgrad v \cdot J^{-1} \cdot \sgrad f,\\
    \partial_t\notvc{u}&= \br{\partial_\tau u_\sigma - v\partial_\sigma u_\sigma + \sigma \sgrad v \cdot J^{-1} \cdot (\sgrad u_\sigma + K u_\bot)}\nhat,\\
    &\quad+\br{\partial_\tau u_\bot - v\partial_\sigma u_\bot + \sigma \sgrad v \cdot J^{-1} \cdot (\sgrad u_\bot - K u_\sigma)}.	
    \end{align*}
\emph{Rescaling} --- Rescaling the normal coordinate by $\xi = \sigma/\eps$ gives the following operators:
	\begin{align*}
    \nabla 	&= \eps^{-1} \nhat \dx + \sum\nolimits_{k=0}^\infty \eps^k \xi^k K^k \sgrad,\\
    |J|\divof\notvc{u}&= \eps^{-1} {\dx u_\sigma}  -\dx(\xi \Kbar u_\sigma) + \sdiv{u_\bot} + \eps \br{\dx(\xi^2 \Kabs u_\sigma) - \xi \sdiv(\Khat u_\bot)},\\
	\lapof f &= \eps^{-2} \dx^2 f + \eps^{-1}(-\Kbar \dx f) + \eps^0(-\xi \overline{K^2}\dx f + \sgrad \cdot \sgrad f) + \O(\eps)\\
    -\nabla\times\nabla\times\notvc{u}&= \eps^{-2} \dx^2 u_\bot + \eps^{-1} \br{- \Kbar \dx u_\bot - \dx \sgrad u_\sigma- \nhat \sgrad \cdot (\dx u_\bot)} + \O(\eps^0)\\
    \notvc{u}\cdot \grad f &= \eps^{-1} u_\sigma \dx f + \sum\nolimits_{k=0}^\infty \eps^k \xi^k K^k u_\bot\cdot \sgrad f\\
	\notvc{u} \cdot \grad\notvc{u}&=\eps^{-1}\left( u_\sigma \dx u_\bot + \nhat u_\sigma\dx u_\sigma\right)\\
	&\quad + \sum\nolimits_{k=0}^{\infty} \eps^k\xi^k u_\bot K^k \cdot \br{(\sgrad u_\bot - K u_\sigma) + (\sgrad u_\sigma+ K u_\bot)\nhat} \nonumber,\\
    \partial_t &= -\eps^{-1} v \dx + \eps^0 \partial_\tau + \eps^1 \left(\xi \sgrad v \cdot\sum\nolimits_{k=0}^\infty\eps^k\xi^k K^k\sgrad\right)
    \end{align*}

\section{Differential geometry of remapped coordinates}
\label{app:remapped-coordinates}
We solve \cref{sec:salty-boussinesq} by remapping to a fixed rectangular domain with coordinates $\tau, \xi,$ and $\zeta^\pm$,
	\begin{align*}
	\tau(t,x,z) &= t, &
	\xi(t,x,z) &= x, &
	\zeta^+(t,x,z) &= \frac{z}{h(t,x)}, &
	\zeta^-(t,x,z) &= \frac{z-h(t,x)}{2-h(t,x)}, &
	\eta(\tau,\xi) &= h(t,x),
	\end{align*}
where $\eta$ is the interface height in the new coordinates.
We now develop the differential geometry required to write \cref{eq:sharp-pdes} and \cref{eq:phase-pdes} in the new coordinates.
We give explicit formulae for the $\zeta^+$ remapping.
The $\zeta^-$ remapping is analogous and straightforward to derive.
We use a tangent vector basis derived from the Jacobian $\J$ of our transformation and its inverse $\K$,
	\begin{align*}
	\J^+ &\equiv 
    \begin{bmatrix}
    \partial_t \tau & \partial_t \xi & \partial_t \zeta \\
    \partial_x \tau & \partial_x \xi & \partial_x \zeta \\
    \partial_z \tau & \partial_z \xi & \partial_z \zeta \\
    \end{bmatrix}
    = \begin{bmatrix}
    {1} & 0 & - \zeta \frac{\partial_\tau \eta}{\eta}\\
    0 & {1}{} & - \zeta \frac{\partial_\xi \eta}{\eta}\\
    0 & 0 & \frac{1}{ \eta}
    \end{bmatrix}, &
	\K^+ &\equiv 
    \begin{bmatrix}
    \partial_\tau t & \partial_\tau x & \partial_\tau z \\
    \partial_\xi t & \partial_\xi x & \partial_\xi z \\
    \partial_{\zeta} t & \partial_{\zeta} x & \partial_{\zeta} z \\
    \end{bmatrix}
    = \begin{bmatrix}
    1 & 0 &  \zeta \partial_\tau \eta\\
    0 & 1 &  \zeta \partial_\xi \eta\\
    0 & 0 &  \eta
    \end{bmatrix}.
	\end{align*}
The tangent vector components are the rows of the spatial component of the inverse Jacobian $e_i = {\K_i}^j \ehat_j$.
We use Einstein notation to sum over the spatial indices $i = 1,2$.
These tangent vectors induce dual vectors $\omega^i = {\J_j}^i \, \ehat^j$ which satisfy $\omega^i \cdot e_j = \delta^i_j$, with components from the column vectors of the spatial component of the Jacobian.
We record the length of the vectors using the metric, with co/contravariant components $g_{ij} = e_i \cdot e_j$, $g^{ij} = \omega^i \cdot \omega^j$,
	\begin{align*}
	\label{eq:coordinate-vectors}
	e_1^+ &= \ehat_1 + \zeta \partial_\xi \eta  \, \ehat_2, &
	e_2^+ &= \eta \,  \ehat_2, &
	{\omega^{1}}^+ &= \ehat_1, &
	{\omega^{2}}^+ &=  \frac{\ehat_2 - \zeta \partial_\xi \eta \, \ehat_1}{\eta},\\
	g_{11}^+ &= (1 + \zeta^2 \partial_\xi \eta^2),  &
	g_{12}^+ &= \zeta \eta \partial_\xi \eta, &
    {g^{11}}^+ &= 1, &
    {g^{12}}^+ &= - \zeta \frac{\partial_\xi \eta}{\eta},\\
	g_{21}^+ &= \zeta \eta \partial_\xi \eta, &
	g_{22}^+ &= \eta^2, &
    {g^{21}}^+ &= - \zeta \frac{\partial_\xi \eta}{\eta}, &
    {g^{22}}^+ &= \frac{1}{\eta^2}\left(1 + \zeta^2 \partial_\xi \eta^2 \right).
	\end{align*}
These give ``Jacobian'' $\|\K\|$ determinants of $\|\K^+\| \equiv \sqrt{\|g_{ij}\|} = \eta$.
The completely antisymmetric tensor can be transformed from Cartesian coordinates,
	\begin{align*}
	\E = [ij]\ohat^i \ohat^j = [ij]\K^i_k \K^j_l \, \omega^k\omega^l = {\|\K\|} [kl]\,\omega^k\omega^l = \frac{1}{\|\K\|}[kl] \, e_k e_l,
	\end{align*}
where $[ij]$ is the antisymmetric symbol.
We project the gradient to the tangent basis with the covariant derivative $\nabla_i = e_i \cdot \nabla \iff \nabla = \omega^i \nabla_i,$.
We measure spatial variation of the basis with the connection coefficients $\Gamma^k_{ij} = \omega^k \cdot \nabla_j e_i \iff e_j \cdot \nabla e_i = \Gamma^k_{ij} e_k$.
For the tangent basis, the connection coefficients take a particularly simple and symmetric form
\begin{align*}
{\Gamma^1_{11}}^+ &= {\Gamma^1_{12}}^+ = {\Gamma^1_{21}}^+ = {\Gamma^1_{22}}^+ = 0, &
{\Gamma^2_{11}}^+ &= \frac{\zeta\dx^2 \eta}{\eta}, & 
{\Gamma^2_{12}}^+ &= {\Gamma^2_{21}}^+ = \frac{\dx \eta}{\eta},&
{\Gamma^2_{22}}^+ &= 0.
 \end{align*}
We note the connection coefficients for the dual vectors are related to those for the tangent basis by $0 = \nabla_k(\omega^i \cdot e_j) = \nabla_k (\omega^i)\cdot e_j + \omega^i \cdot \nabla_k (e_j) = \nabla_k (\omega^i)\cdot e_j + \Gamma^i_{jk} $.
The partial time derivative $\partial_t$ changes as defined in zeroth row of the Jacobian, and the basis vectors also evolve in time,
    \begin{align*}
    \partial_\tau e_1^+ &\equiv {\Gamma^i_{10}}^+ e_i^+ = \frac{\zeta}{\eta}\partial_\tau\partial_\xi \eta\, e_2^+, &
    \partial_\tau e_2^+ &\equiv {\Gamma^i_{20}}^+ e_i^+ = \frac{\partial_\tau\eta}{\eta}\, e_2^+.
    \end{align*}
These geometric quantities allow us to calculate all the relevant vector calculus quantities,
	\begin{align*}
	\nabla\notvc{u}&= \nabla_i(u^j \, e_j)\omega^i = (\nabla_i u^j + u^k\Gamma^j_{ki})\,e_j \, \omega^i,\\
	\nabla \cdot\notvc{u}&= \nabla\notvc{u}: I = (\nabla_i u^j + u^k\Gamma^j_{ki})\,e_j \cdot \omega^i = \nabla_i u^i + u^k \Gamma^i_{ki},\\
	\nabla \times\notvc{u}&= \mathcal{E}^\top:\nabla\notvc{u}= (\mathcal{E}_{ij}\omega^i \omega^j):\omega^k(\nabla_k)(u^l \, e_l) = \|\K\| [ij] u^l_{;k}(\omega^i\cdot \omega^k)(\omega^j\cdot e_l) = \|\K\| [ij] g^{ik} u^j_{;k} ,\\
	\nabla \times q &= \mathcal{E}^\top\cdot\nabla q = (\mathcal{E}^{ij} \nabla_k q)(e_j\cdot\omega^k)e_i   =  \frac{1}{\|\K\|}[ij]q_{,j} e_i = \frac{1}{\|\K\|}(q_{,2}e_1 - q_{,1}e_2) ,\\
	\nabla^2 f &= \nabla_i( \nabla_j f \,\omega^j)\,\omega^i : I = (\nabla_i\nabla_j f )\, \omega^j \cdot \omega^i - \nabla_j f\Gamma^j_{ki}\omega^k \cdot \omega^i = g^{ji}\nabla_i\nabla_j f - g^{ki}\nabla_j f \Gamma^j_{ki},\\
	\nabla p &= g^{ji} p_{,j} e_i,\\
    u\times q &= \mathcal{E}\cdot\notvc{u}q = (\mathcal{E}_{ij}\omega^i\omega^j)\cdot(q u^k e_k) = \|\K\| g^{jk} [ij] u^i e_k = \|\K\|q(g^{2k} u^1 - g^{1k}u^2)e_k,\\
	\nabla\times\nabla \times\notvc{u}&= [ij]\partial_j([kl]\partial_k u_l) \, \ehat_i=(\delta_{ik}\delta_{jl}-\delta_{il}\delta_{jk})\partial_j\partial_k u_l \ehat_i = \nabla(\nabla\cdot u) - \nabla^2 u, \\
    (\nabla \times u)\times\notvc{u}&= [ij][kl]\partial_k u_l u_j \ehat_i = (\partial_i u_j-\partial_j u_i) u_j \ehat_i = \tfrac{1}{2}\nabla(u\cdot u) - (u\cdot \nabla )u,\\
    \nabla \cdot (u u) &= (u\cdot \nabla)\notvc{u}+ (\nabla \cdot u) u,\\
	\partial_t &= \partial_\tau + {\J_0}^i \nabla_i,\\
	\partial_t\notvc{u}&= \partial_t u^i e_i + u^i \partial_t e_i =  (\partial_\tau + \J_0^{i}\nabla_i) u^j e_j + {\J_0}^{i\geq0}\Gamma^{j}_{ki\geq0} u^k e_j.
	\end{align*}
The normal vector at the interface is proportional to the $\zeta$ dual vector.
This gives us the normal gradient and normal velocity at the interface,
    \begin{align*}
    \nhat &= \frac{\omega^2}{\sqrt{g^{22}}} &
    \nhat \cdot \nabla &= \frac{g^{2i}}{\sqrt{g^{22}}}\nabla_i, & 
    v &= \partial_t h \, \ehat_z, &
    \nhat \cdot v &= \frac{1}{\sqrt{g^{22}}}\frac{\partial_t h}{h}.
    \end{align*}
We finally write the interfacial curvature as $	\kappa = {\partial_\xi^2 \eta}/{(1 + \partial_\xi \eta^2)^{3/2}}$.
This completes the relations used to simulate the sharp interface and phase-field equations in remapped geometries.
\bibliographystyle{siam}
\bibliography{phase-field-references.bib}

\end{document}